\documentclass[aps,prb,reprint,longbibliography,superscriptaddress]{revtex4-2}
\usepackage{mathrsfs}
\usepackage{amsmath,gensymb}
\usepackage{amsfonts}
\usepackage{amssymb}
\usepackage{amsthm}
\usepackage{graphicx}
\usepackage{natbib}
\usepackage{color}
\usepackage{hyperref}
\usepackage{bm}
\usepackage[caption=false]{subfig}
\usepackage{verbatim}
\usepackage[normalem]{ulem}

\begin{document}

\title{Voltage-driven dynamics of $\varphi_0$-S/F/S Josephson junctions chains}

\author{G. A. Bobkov}
\affiliation{Moscow Institute of Physics and Technology, Dolgoprudny, 141700 Moscow region, Russia}

\author{A. M. Bobkov}
\affiliation{Moscow Institute of Physics and Technology, Dolgoprudny, 141700 Moscow region, Russia}

\author{I.V. Bobkova}
\affiliation{Moscow Institute of Physics and Technology, Dolgoprudny, 141700 Moscow region, Russia}
\affiliation{National Research University Higher School of Economics, 101000 Moscow, Russia}

\begin{abstract}
Superconductor/ferromagnet/superconductor Josephson junctions with anomalous phase shift $\varphi_0$ ($\varphi_0$-S/F/S JJs) implement  a coupling between the superconducting phase and the spin degrees of freedom. Here we investigate the  dynamics of voltage-biased coupled chains of $\varphi_0$-S/F/S JJs and predict that the presence of $\varphi_0$ makes the conventional regime corresponding to the linear growth of the superconducting phase at each of the junctions and oscillating Josephson current unstable. New stable regimes are found and investigated. The changes of the dynamic behavior are clearly seen in the IV-characteristics of the system and can serve as a fingerprint of the presence of the magnetoelectric coupling $\varphi_0$. Moreover, the collective magnetic excitations of the chains of $\varphi_0$-S/F/S JJs, which were reported by G.A. Bobkov {\it et. al.} [JETP Lett. {\bf 119}, 251 (2024)] also manifest themselves in the IV-characteristics. This provides a method for their experimental detection and direct measurement of the constant quantifying the strength of the  coupling between
the superconducting phase and the spin degrees of freedom.
\end{abstract}

\maketitle

\section{Introduction}

Superconductivity is a macroscopic quantum state characterized as a whole by a superconducting phase. A basic property of superconducting systems is that the charge current and the superconducting phase are directly connected. A canonical example of such a relationship is a current-phase relationship (CPR) of a Josephson junction (JJ) \cite{Golubov_review}. The minimal form of the CPR, corresponding to a single harmonic contribution,
is given by $j(\varphi) = j_c \sin \varphi$. Here $|j_c|$ is the critical current
and $\varphi$ is the phase difference between superconducting electrodes. The ordinary Josephson junctions have
$j_c > 0$ yielding the zero phase difference ground state $\varphi=0$. In certain cases $j_c < 0$ leading to the ground state $\varphi=\pi$. Such $\pi$-junctions have been realized in many systems, including  superconductor/ferromagnet/superconductor (S/F/S) JJs \cite{Buzdin1982,Buzdin2005,Ryazanov2001,Kontos2002,robinson2006critical},
non-equilibrium  superconductor/normal metal/superconductor (S/N/S) JJs \cite{Baselmans1999,golikova2021controllable} and many others \cite{schulz2000design,jorgensen2007critical,vanDam2006,ke2019ballistic}. They can be used in superconducting logic and quantum computers\cite{Feofanov2010,Yamashita2005,Shcherbakova2015}.

Under the simultaneous breaking of time-reversal and inversion symmetries even more exotic situation can be realized in JJs. The CPR can take the form $j(\varphi) = j_c \sin (\varphi - \varphi_0)$ with the anomalous (spontaneous) ground state phase shift $\varphi_0 \neq 0,\pi$ \cite{Bobkova_review,Shukrinov_review}. Such JJs have already been implemented experimentally in systems with spin-orbit coupling, which is a manifestation of the inversion symmetry breaking, under the action of the applied magnetic field, which breaks the time-reversal symmetry \cite{Mayer2020,Szombati2016,Assouline2019,Murani2017}.

Another possibility to break the time-reversal symmetry is to use JJs via ferromagnetic interlayers (S/F/S JJs). This is more interesting because the anomalous ground state phase shift $\varphi_0$ depends on the direction of the interlayer magnetization and thus provides a direct magnetoelectric coupling between the magnetization and the superconducting phase. Therefore, $\varphi_0$-S/F/S JJs implement not only a canonical coupling between the electric current and the phase, but also a coupling between the phase and the spin degrees of freedom. Because of this, such structures look very promising for applications in superconducting spintronics \cite{Linder2015,Eschrig2015}, especially for magnetization control \cite{Konschelle2009,Shukrinov2017,Nashaat2019,Rabinovich2019,Guarcello2020,Bobkova2020,Bobkova_review}. The choice of systems in which the spin-orbit interaction can be strong enough and, accordingly, such a coupling between the phase and the spin can be realized, is quite wide. For example, hybrids of a ferromagnetic insulator/3D topological insulator can be used as interlayers \cite{Chang2013,Kou2013,Kou2013_2,Chang2015,Jiang2014,Wei2013,Jiang2015,Jiang2016}. It is also possible to use thin-film materials in which there is a structural inversion symmetry breaking, for example, 2D or quasi-2D ferromagnets \cite{Gibertini2019,Ai2021,Kang2022}.

Recently it was predicted that, besides controlling the magnetization of a single ferromagnet, which is a weak link of the $\varphi_0$-S/F/S JJ, the direct coupling between the phase and the magnetization provides a possibility to establish and control a collective behavior of magnetic moments of different weak links in chains of $\varphi_0$-S/F/S JJs \cite{Bobkov2022,Bobkov2024modes,Bobkov2024many}. The collective effects occur due to the extremely long-range interaction between the magnetic moments mediated by the macroscopic superconducting phase \cite{Bobkov2022}.

Here we investigate the behavior of such coupled chains of $\varphi_0$-S/F/S JJs under the applied external dc voltage $V$ and demonstrate that the presence of the coupling between the phase and the magnetic moment clearly manifests itself in the IV-characteristics of the system. Nonzero $\varphi_0$ makes the conventional regime, which corresponds to the linear growth of the superconducting phase at each of the junctions and oscillating Josephson current and usually occurs in the coupled chains of JJs, unstable. The ac electric current induced by the applied voltage excites collective magnetic oscillations in such a system. Due to the magnetoelectric coupling they cause a back action on the superconducting phase, which can result in the development of instability.  The instability begins to increase at voltages that exceed an eigenfrequency of the system \cite{Bobkov2024modes} and the chain goes into another stable regime, in which the superconducting phase differences at individual JJs are not linear functions of time. This is accompanied by a significant freezing of the amplitude of oscillations of the magnetic moments and electric current in the chain. In addition the oscillation frequency of the current and the magnetic precession becomes a multiple integer of the Josephson frequency. This dynamic regime occurs in a wide range of voltages exceeding the voltages corresponding to eigenfrequencies of the system. Thus, it is very robust and can serve as a fingerprint of the presence of the magnetoelectric coupling $\varphi_0$. While the proposed and implemented in the literature methods to detect the anomalous phase shift are based on the SQUID technology, we propose a simple, easily accessible way to detect the presence of the anomalous phase shift and magnetization-phase coupling in the chains of Josephson junctions by electrical means. It is also shown that there is another stable dynamical regime, which is characterized by beats in the Josephson current with a period that corresponds to the difference in frequencies between the acoustic and optical modes of the collective oscillations of the chain \cite{Bobkov2024modes}. This regime is also clearly seen at the IV-characteristic of the chain, which provides a method of an experimental detection of the collective magnetic excitations.
Moreover, the difference between the eigenfrequencies can be found from the beats of ac
electric current. This allows for a direct
measurement of the constant, quantifying the strength of the magnetoelectric coupling between the superconducting phase and the magnetic moment.

\section{Model and method}
\label{model}

\begin{figure}[tb]
	\begin{center}		\includegraphics[width=70mm]{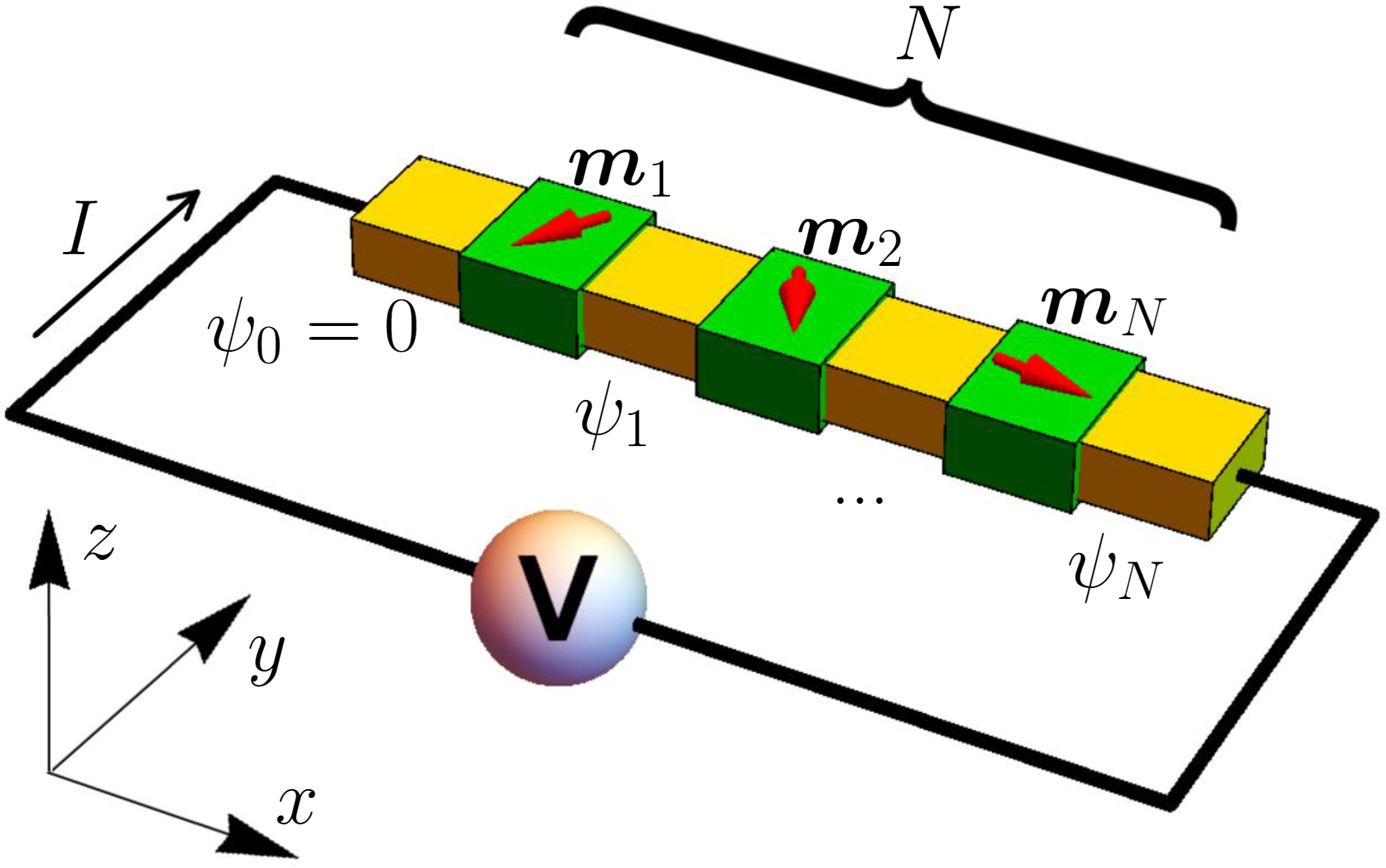}
		\caption{Sketch of a coupled system of $N$ $\varphi_0$-S/F/S JJs. The dc voltage $V$ is applied between the external leads. Magnetic moment of each JJ is shown by a red arrow. $\psi_i$ is a phase of the superconducting region connecting $i$-th and $i+1$-th weak links.}
  \label{fig:sketch}
	\end{center}
 \end{figure}

We consider a linear chain of $N$ coupled $\varphi_0$ - S/F/S JJs, where S means a conventional superconductor and F means a homogeneous ferromagnet characterized by a unit vector $\bm m_i = \bm M_i/|\bm M_i|$ along the direction of the magnetic moment $\bm M_i$ of the $i$-th magnet in the chain, see Fig.~\ref{fig:sketch}. It is assumed that the ferromagnets are easy-axis magnets with the easy-axis  along the $z$-direction. The CPR of an individual S/F/S JJ takes the form $I = I_{c,i} \sin (\varphi_i-\varphi_{0,i})$, where $I_{c,i}$ is the critical current of $i$-th JJ, $\varphi_i = \psi_i - \psi_{i-1}$ is the superconducting phase difference at this JJ and $\varphi_{0,i}$ is the anomalous phase shift for a given JJ. $\varphi_0$ determines the form of the magnetoelectric coupling between the superconducting phase and the magnetic moment $\bm m_i$. The specific dependence of $\varphi_0$ on the direction of magnetization is determined by the type of spin-orbit interaction (SOC) that is present in the system. Low-dimensional ferromagnets and thin-film hybrid structures ferromagnet/normal metal (F/N) are characterized by structural inversion asymmetry and thus the most common type of SOC in such systems is the Rashba SOC. In this case the anomalous phase shift takes the form 
\begin{eqnarray}
\varphi_{0,i} =  r_i \hat {\bm  j} \cdot (\bm n \times \bm m_i ),
\label{phi_0}
\end{eqnarray}
where $\hat {\bm  j}$ and $\bm n$ are unit vectors along the Josephson current and along the direction of the structural asymmetry in the weak link, respectively. We consider a thin-film geometry of the weak link with $n$ along the $z$-axis, which is perpendicular to the Josephson current flowing along the $x$-axis. The strength of the coupling between magnetization of the $i$-th ferromagnet and superconducting phase depends on the magnetization of the ferromagnet,  impurity concentration in it, the strength of the SOC, length and geometry of the ferromagnetic interlayer and can be described by a constant $r_i$. A number of theoretical studies calculated  this constant for different model systems, such as ballistic and diffusive metallic ferromagnets with  Rashba SOC or S/F/S JJs on top of the 3D TI \cite{Buzdin2008,Bergeret2015,Zyuzin2016,Nashaat2019}. 

However, the results presented below depend only on the symmetry of Eq.~(\ref{phi_0}) expressing how the anomalous phase shift depends on the direction of the magnetization $\bm m_i$. For the case under consideration this symmetry dictates that
\begin{eqnarray}
\varphi_{0,i} = r_i m_{yi}.
\label{chi0}
\end{eqnarray}
The dependence of the critical current $I_{c,i}$ on the direction of $\bm m_i$ depends on the particular type of the considered S/F/S JJ. It can be independent on the magnetization direction, as it has been reported for the ferromagnetic nanowires with SOC \cite{Buzdin2008}, or it can depend strongly on  the $x$-component of the magnetization, as it takes place for the ferromagnetic interlayers on top of the 3D TI \cite{Zyuzin2016,Nashaat2019}.  In the present manuscript we focus on the class of systems for which the direction-independent critical current was reported because they represent the simplest and simultaneously realistic objects for studying the dynamic phenomena produced by the magnetoelectric coupling $\varphi_0$. The influence of the dependence $I_{c,i}(\bm m_i)$ on the behavior of the voltage-driven $\varphi_0$-S/F/S chain is a prospect for future work.

The voltage-driven dynamics of the system can be calculated on the basis of two coupled equations. The dynamics of magnetic moments $\bm m_i$ is described by the Landau-Lifshitz-Gilbert (LLG) equation:
\begin{eqnarray}
\frac{d \bm m_i}{d t} = -\gamma \bm m_i \times \bm H_{M}^i + \alpha_i \bm m_i \times \frac{\partial\bm m_i}{\partial t} - \nonumber \\
\frac{\gamma r_i \hbar I (t)}{2e M V_F}[\bm m_i \times \bm e_y],~~~~~~
\label{LLG}
\end{eqnarray}
where $\gamma$ is the gyromagnetic ratio,  $\bm H_{M}^i $ is the local effective field in the ferromagnet induced by the easy-axis magnetic anisotropy. For simplicity all the JJs are assumed to be the same, that is  $\bm H_{M}^i \equiv (K/M) m_{zi} \bm e_z$, where $K$ is the anisotropy constant, $M$ is the magnetization of a magnet and $V_F$ is its volume, thus $E_M^0 = KV_F/2$ is the anisotropy energy of the magnet. Contributions to $\bm H_M^i$ resulting from the dipole-dipole interactions with other magnets in the chain can be safely disregarded because the phase-mediated interaction between the magnets is very long-range and does not decay even at submillimeter scales \cite{Bobkov2022}. For typical parameters of the JJs with 3D topological insulator/ferromagnetic insulator interlayers \cite{Veldhorst2012,Xiao2010} at such distances between the magnets the energy of the dipole-dipole interaction should be several orders of magnitude smaller than the Josephson energy, which accounts for the phase-mediated interaction. $\alpha_i = \alpha$ is the Gilbert damping constant, $r_i = r$ and $I_{c,i} = I_c$. Relatively weak inhomogeneity of JJs parameters does not lead to qualitative changes in the results, see Appendix \ref{variations}. The last term in Eq.~(\ref{LLG}) describes the spin-orbit torque, exerted on the magnet by the electric current $I (t)$  \cite{Yokoyama2011,Miron2010,Bobkova2018,Bobkova2020,Bobkov2022}. 

The total current flowing through each of the JJs can be calculated in the framework of the RSJ-model and consists of the supercurrent and the normal quasiparticle current contributions. In the case of a $\varphi_0$-JJ the voltage drop across the $i$-th junction $V_i$ 
is related to the time derivative of the phase difference at this JJ as $d(\varphi_i)/dt = 2eV_i/\hbar$. However, in this case the equivalent circuit scheme of the junction contains not only an ideal JJ with zero resistance and a normal resistance of the JJs connected parallel, but also a "battery" producing the electomotive force from the emergent electric field $\propto \dot \varphi_0$ \cite{Rabinovich2019}. For this reason the normal quasiparticle current contribution, which is determined by the voltage drop at the normal resistance of the JJ, is related to the gauge invariant phase difference $\varphi_i - \varphi_{0,i}$ \cite{Rabinovich2019}: 
\begin{eqnarray}
I(t)=I_c \sin (\varphi_i - \varphi_{0,i})+ 
\frac{\hbar}{2eR_N}\frac{d({\varphi}_i  -  {\varphi}_{0,i})}{dt},
\label{current_total}
\end{eqnarray}
where $R_N$ is the normal state resistance of an individual S/F/S JJ. Here we consider metallic weak links and  disregard the capacitance of the JJs. In the most common model for implementation of the voltage-biased regime it is assumed that the voltage $V$ is produced by a generator consisting of an ideal voltage supply $V_g$ and internal resistance $r_g$. Under the conditions $r_g/R_N \ll 1$ and $I_c r_g \ll V_g$ the voltage bias regime is feasible and $V \approx V_g$.  With the condition of a dc applied voltage $V$, which gives $\psi_N = 2eVt/\hbar$, Eqs.~(\ref{LLG}) and (\ref{current_total}) allow for determination of all the  dynamical quantities describing the behavior of the system: the phase differences $\varphi_i(t) = \psi_i(t)-\psi_{i-1}(t)$, the magnetic moments of all the links $\bm m_i(t)$ and the current $I(t)$. The details of the numerical procedure are described in Appendix \ref{numerical}.

\begin{figure}[tb]
	\begin{center}		\includegraphics[width=70mm]{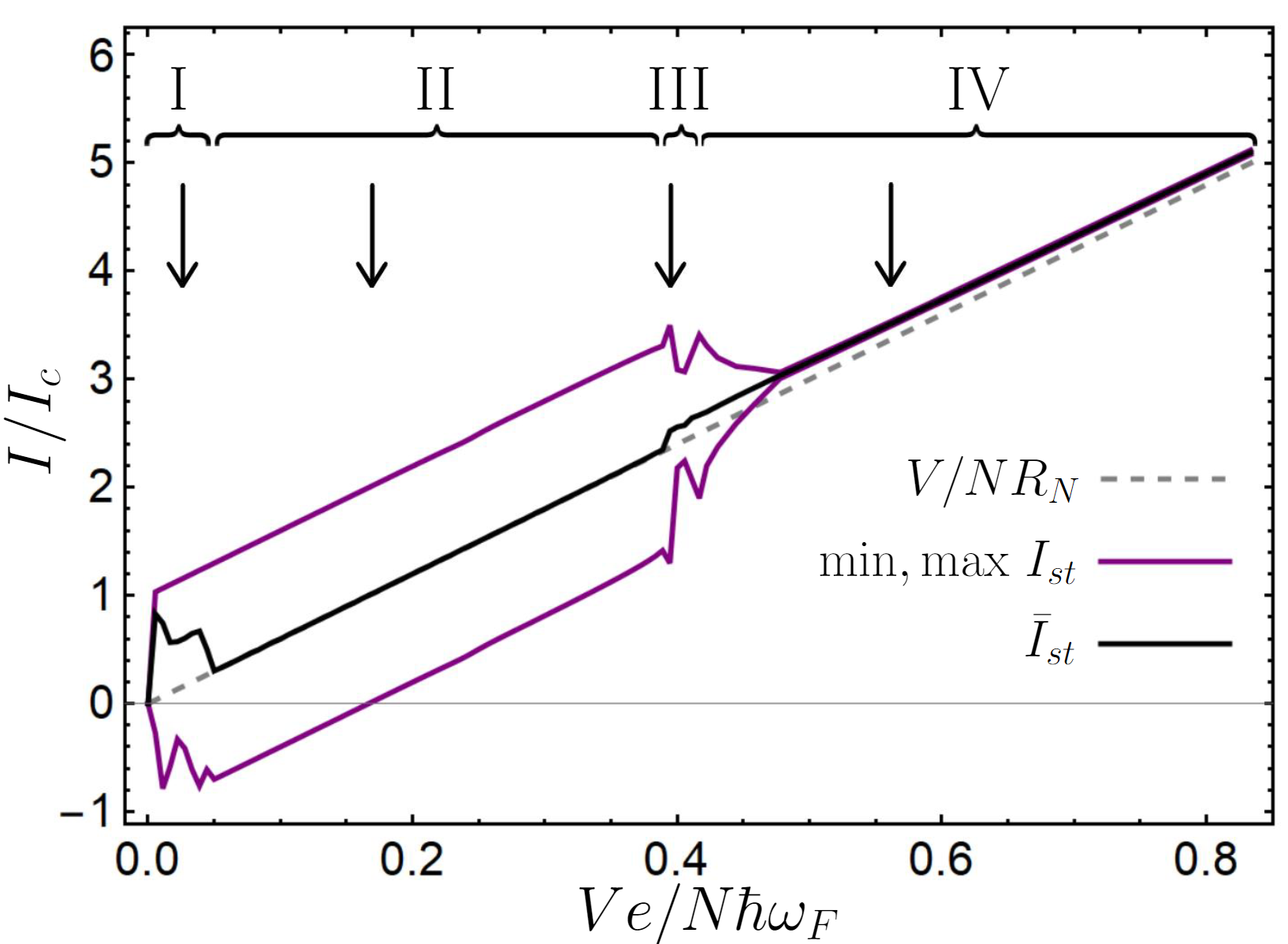}
		\caption{IV-characteristics (solid black line) of the $\varphi_0$-S/F/S JJs chain under the applied dc voltage. Amplitude of the current oscillations is shown by purple lines. The normal Ohm's law is shown by the dashed line. Different regimes of the voltage-driven dynamics are marked by numbers I-IV. $N=3$, $E_{J}^0/E_{M}^0=2/3$, $r=0.3$, $I_cR_Ne=0.17 \hbar \omega_F$, $\alpha=0.02$. Black arrows indicate particular voltages at which the dynamics of the system is demonstrated below.}
  \label{fig:IVcharacteristics}
	\end{center}
 \end{figure}

There is an important physical reason to consider the regime of applied dc voltage, and not of applied dc current. The point is that we are interested in the signatures of the {\it collective} behavior of the magnetic weak links, which is a result of their interaction established via the superconducting phase and controlled by the external phase difference \cite{Bobkov2022}. The regime of applied current is also frequently studied, but in relation to the system under consideration according to Eq.~(\ref{LLG}) the applied current generates independent dynamics of all the JJs in the chain. Although it is also an important problem and some interesting effects including a chaotic behavior have already been predicted \cite{Mazanik2024}, the collective behavior of the magnets is not excited in the applied current regime.

\begin{figure*}[tb]
	\begin{center}		\includegraphics[width=165mm]{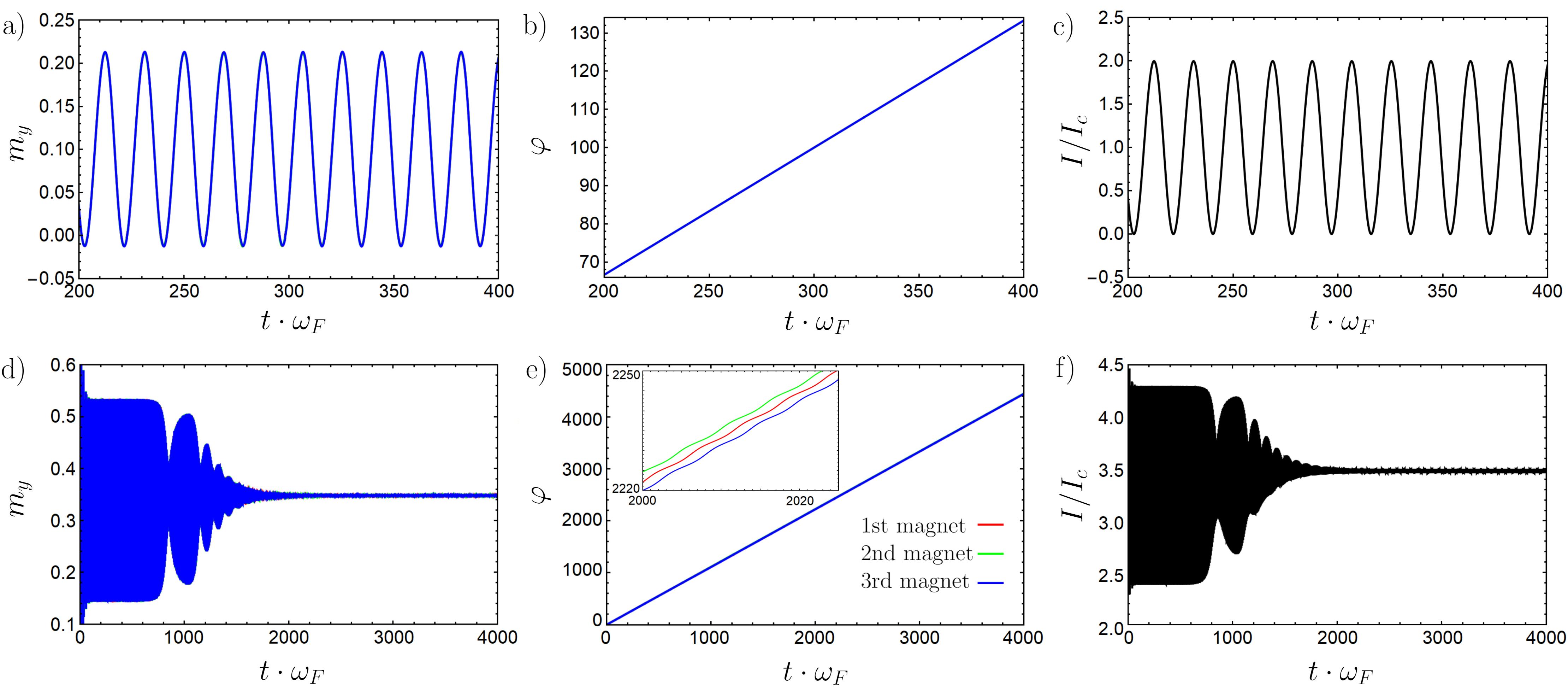}
		\caption{Dynamics of $\varphi_0$-S/F/S JJs chain with $N=3$. Upper row: regime II. (a) $m_{yi}(t)$ [the same for all JJs]; (b) $\varphi_i(t)$ [the same for all JJs] and (c) $I_{st}(t)$ at $eV=0.5\hbar \omega_F$. Only the stationary state is shown. Bottom row: regime IV. (d) $m_{yi}(t)$ [the same for all JJs]; (e) $\varphi_i(t)$, which are different for all magnets, which is shown on a larger scale in the inset. Here and below red, green and blue curves correspond to the 1st, 2nd and 3rd magnets, respectively. (f) $I(t)$ at $eV=1.67 \hbar \omega_F$. The full time range is shown including the initial and transitional regimes. }
  \label{fig:regimes}
	\end{center}
 \end{figure*}

Because we are interested in the collective behavior of the system of coupled magnets and Josephson phases; first, let us briefly recall what the collective modes of magnetic excitations induced by the phase-mediated interaction between magnets in such a system look like \cite{Bobkov2024modes}. At $V=0$ and, consequently, $\psi_N=0$, the equilibrium states of all the magnets are along the easy $z$-axis. In Ref.~\onlinecite{Bobkov2024modes} it was found that such a chain of coupled $\varphi_0$-S/F/S JJs has two eigenmodes of small excitations with the frequencies:
\begin{equation}
\omega_a = \gamma \sqrt{\frac{K}{M}(\frac{K}{M} -   \lambda)}, 
\label{frequency_1_z}
\end{equation}
\begin{equation}
\omega_o = \gamma \frac{K}{M}, 
\label{frequency_2_z}
\end{equation}
where $\lambda = -r^2E_J^0/V_F M$ with $E_J^0 = \hbar I_c/2e$ being the amplitude of the Josephson energy of a JJ. The small corrections to the frequencies due to the Gilbert damping constant $\alpha$ are neglected. $\omega_a$ is an acoustic mode because it corresponds to the motion of all the magnetic moments $\delta \bm m_i(t) = (m_{xi}(t), m_{yi}(t),0)$ with the same phase.  $\omega_o$ can be interpreted as an optic frequency corresponding to $N-1$ different modes with  $\sum \limits_i m_{yi}(t) = 0$. When the equilibrium position of the magnetic moments is along the $z$-axis $\omega_o$ coincides with the frequency of the ferromagnetic resonance of an isolated magnet $\omega_F = \gamma (K/M)$, but in general it may be not the case \cite{Bobkov2024modes}.  At $V \neq 0$ both frequencies depend on the external phase $\psi_N$\cite{Bobkov2024modes}.  Nevertheless, the general conclusion that the system possesses acoustic and optical eigenmodes remains valid and the difference $\omega_a - \omega_o$ remains approximately constant and manifests itself in the dynamics of the system, as we will see below.

\section{Different regimes of voltage-driven dynamics}
\label{regimes}

Now we turn to the discussion of dynamical results obtained by solving Eqs.~(\ref{LLG}) and (\ref{current_total}). We apply dc voltage $V$ to the chain at $t=0$ and numerically study the dynamics at $t>0$. After the voltage is turned on, the system undergoes a process of establishing a steady state for some time. In the steady state, the Josephson current and the magnetic moments $\bm m_i$ are periodic functions of time (or a sum of periodic functions). The Josephson current in the steady state is denoted by $I_{st}(t)$. Fig.~\ref{fig:IVcharacteristics} demonstrates the IV-characteristics of the system. The black line is averaged over the period of oscillations value of the current $\bar I_{st}$. The purple lines demonstrate the maximal and minimal values of $I_{st}(t)$. We observe that depending on $V$ the chain can be in 4 essentially different regimes, which are clearly seen in the IV-characteristics and are marked by the corresponding numbers.

In low-voltage regime I the dynamics of the phase differences $\varphi_i (t)$ is determined by frequent phase slips, which results in the appearance of a typical nonmonotonic excess current at the IV-characteristics. This regime is analogous to the voltage-driven regime of  superconducting nanowires\cite{Vodolazov2003}.  It is not related to the magnetoelectric coupling between the phase and the magnetization and can be implemented even in coupled chains of conventional JJs with $\varphi_0 \equiv 0$. For this reason the detailed discussion of regime I is provided in  Appendix \ref{low-voltage}.

The ``regular'' regime II is the most expected. The dynamics of $\varphi_i(t)$, $I_{st}(t)$ and $m_{yi}(t)$ in this regime is presented in the upper row of Fig.~\ref{fig:regimes}. The phase difference is distributed uniformly among all the JJs and grows linearly $\varphi_i(t) = 2eVt/\hbar N$.  $I_{st}(t)$ experiences oscillations with the Josephson frequency $\omega_J = 2eV/\hbar N$. According to Eq.~(\ref{LLG}) it acts on magnetic moments like an oscillating effective field along the $y$-direction, thus exciting their precession with the same frequency around the equilibrium direction, which is very close to the $z$-axis, but is slightly inclined from that direction due to nonzero $\bar I_{st}$ \cite{Konschelle2009,Bobkov2024modes}. 

We obtained that in the absence of the magnetoelectric coupling $\varphi_0 \equiv 0$ regime II is stable for all voltages higher than the boundary of regime I. However, in the presence of the magnetoelectric coupling it appears to be unstable at voltages, that correspond to $\omega_J$ exceeding an eigenfrequency of the system. The existence of such an instability can be found analytically. To do this let us assume that $\varphi_i(t) = \omega_J t + \delta \varphi_i$ with $|\delta \varphi_i| \ll 1$. Then the electric current is $I(t) = I_c \sin (\omega_J t + \delta \varphi_i - r m_{yi})+ (1/2eR_N)(\omega_J + \delta \dot \varphi_i - r \dot m_{yi})$. This oscillating current acts as an exciting force on the magnets.  The oscillations of the magnetic moments are also assumed to be small $m_{yi}(t) \equiv m_y(t) = m_0 + \delta m \sin(\omega_J t - r m_{y} -\beta)$, where $\beta$ is a possible phase shift between the exciting force and the oscillations of the magnets. Up to the first order with respect to $\delta \varphi_i$ we have $I(t) = I_0(t) + \delta I(t)$ with $I_0(t) = I_c \sin (\omega_J t - r m_{y})+ (1/2e R_N)(\omega_J  - r \dot m_{y})$ and $\delta I(t) = I_c \cos (\omega_J t - r m_{y}) \delta \varphi_i + (1/2e R_N) \delta \dot \varphi_i$. Due to the external condition $\psi_N = N\omega_J t$ we have $\sum \limits_{i=1}^N \delta \varphi_i = 0$ which leads to $\delta I(t) = 0$. Linearizing this differential equation also with respect to $\delta m$ we obtain: 
\begin{eqnarray}
\frac{\delta \dot \varphi_i}{2e R_N I_c}+\bigl[\cos (\omega_J t-rm_0) +  ~~~~~~~~~~~~~~\bigr.  \\
\bigl .  + r \delta m \sin (\omega_J t -rm_0 - \beta)\sin( \omega_J t-rm_0) \bigr]\delta \varphi_i= 0 .\nonumber
\label{eq_stability}
\end{eqnarray}
The solution of this equation takes the form:
\begin{eqnarray}
\delta \varphi_i \propto e^{\displaystyle -\kappa t' + \frac{2eR_N I_c}{\omega_J}(\frac{r \delta m\sin(2\omega_J t' - \beta)}{4}-\sin \omega_J t')}
\label{sol_stability}
\end{eqnarray} 
where $\kappa = reR_N I_c \delta m \cos \beta$ and $t'=t-rm_0/\omega_J$. The solution $\varphi_i(t) = \omega_J t$ becomes unstable only in the presence of the magnetoelectric coupling $r$ and when $\kappa < 0$, that is $\beta > \pi/2$, which happens if $\omega_J$ exceeds an eigenfrequency of the system. The simplest analytical estimates can be obtained under the condition of small $r$. Then, from Eqs.~(\ref{current_total}) and (\ref{eq:A3}) it follows that up to leading order in $r$ the ac field takes the form $H_{ac,y} = [r\hbar/(2eMV_F)]I_c \sin \omega_J t$. Then standard linearization of the LLG equation with respect to the ac field and oscillating components of the magnetization gives $m_y = m_0 + \gamma  \omega_F  H_{ac,y} \sin (\omega_J t)/(\omega_F^2 - \omega_J^2) $. It is seen that at $\omega_J >\omega_F$ $\beta = \pi$, as it should  according to the general resonance theory.

\begin{figure}[tb]
	\begin{center}		\includegraphics[width=70mm]{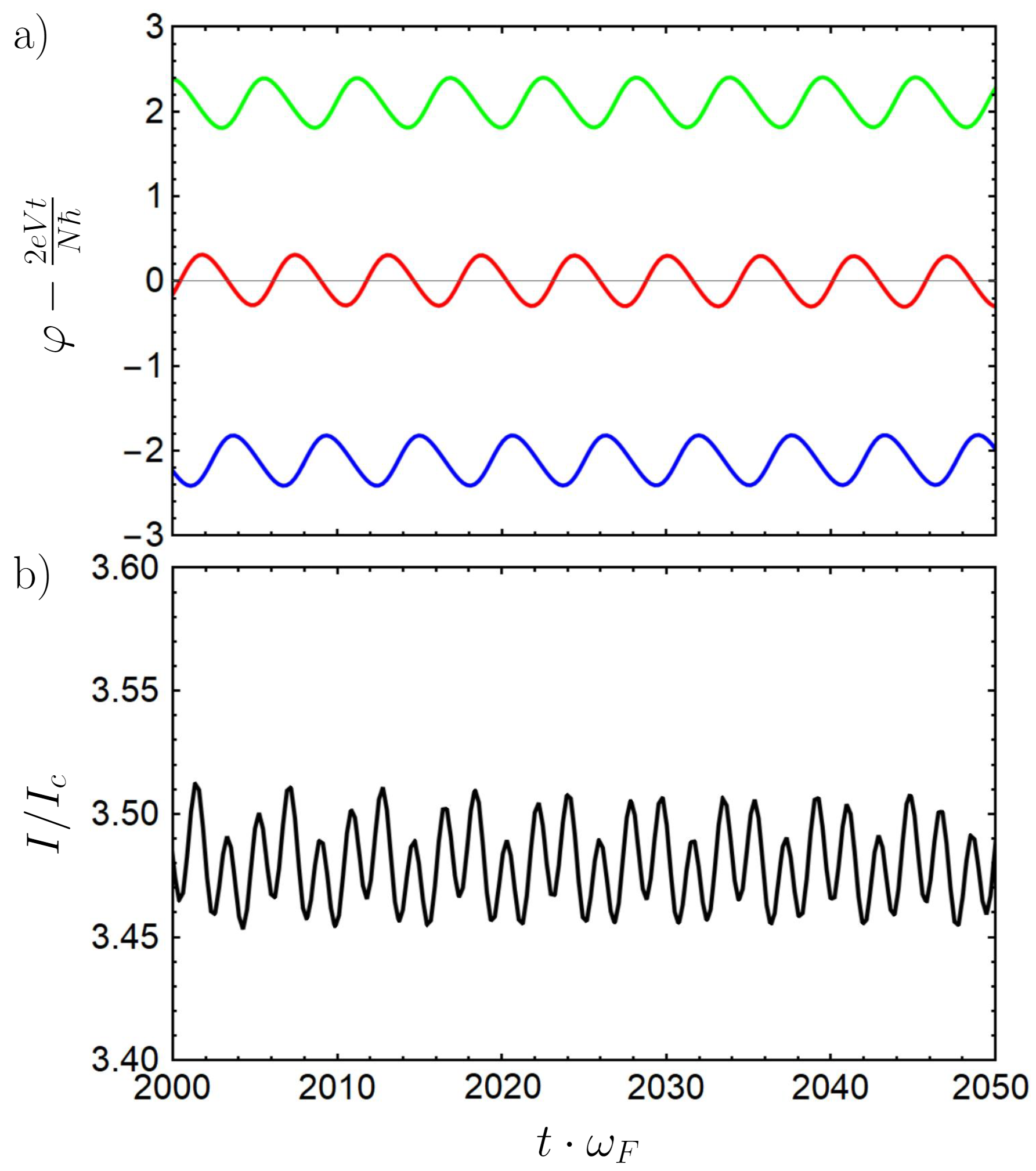}
		\caption{Regime IV. Dynamics of (a) $\varphi_i(t)$ (with subtraction of the linear growth) and (b) $I_{st}$(t) after establishing the steady state. $eV=1.67 \hbar \omega_F$. }
  \label{fig:regimeIV}
	\end{center}
 \end{figure}

For higher voltages, when Regime II becomes unstable the system goes to other stable regimes III and IV. The dynamics of $\varphi_i(t)$, $I_{st}(t)$ and $m_{yi}(t)$ in regime IV is shown in the bottom row of Fig.~\ref{fig:regimes}. At small $t$, the system behaves like in regime II, but after a quite long process of increasing phase instability and transient process, which are described in more detail in Appendix \ref{instability}, it goes to a new stable regime. In this regime the linear in time growth of the phase $\varphi_i(t)$ is accompanied by oscillations with the frequency $\omega_J$ with constant phase shifts between different JJs. This behavior is shown in the inset to Fig.~\ref{fig:regimes}(e) and in more detail in Fig.~\ref{fig:regimeIV} together with the behavior of $I_{st}(t)$. The current $I(t)$ depends on  phase differences $\varphi_i$ at all JJs. Due to the oscillations of the phases $\varphi_i(t)$ the frequency of $I_{st}(t)$ becomes a multiple integer of $\omega_J$. The leading contribution to the current is given by the frequencies, which do not exceed $N\omega_J$. The amplitude of oscillations of the current and the magnetic moments becomes considerably smaller than in regime II and, simultaneously, the averaged value of the current 
$\bar I_{st}$ increases, which is clearly seen in the IV-characteristics.  It is worth noting that at $\varphi_0 = 0$ regime IV is also stable. However, we were able to numerically observe it only by turning off $\varphi_0$ at some $t$. If $\varphi_0=0$ from the very beginning, we were not able to switch the chain from regime II to regime IV. 

\begin{figure}[tb]
	\begin{center}		\includegraphics[width=70mm]{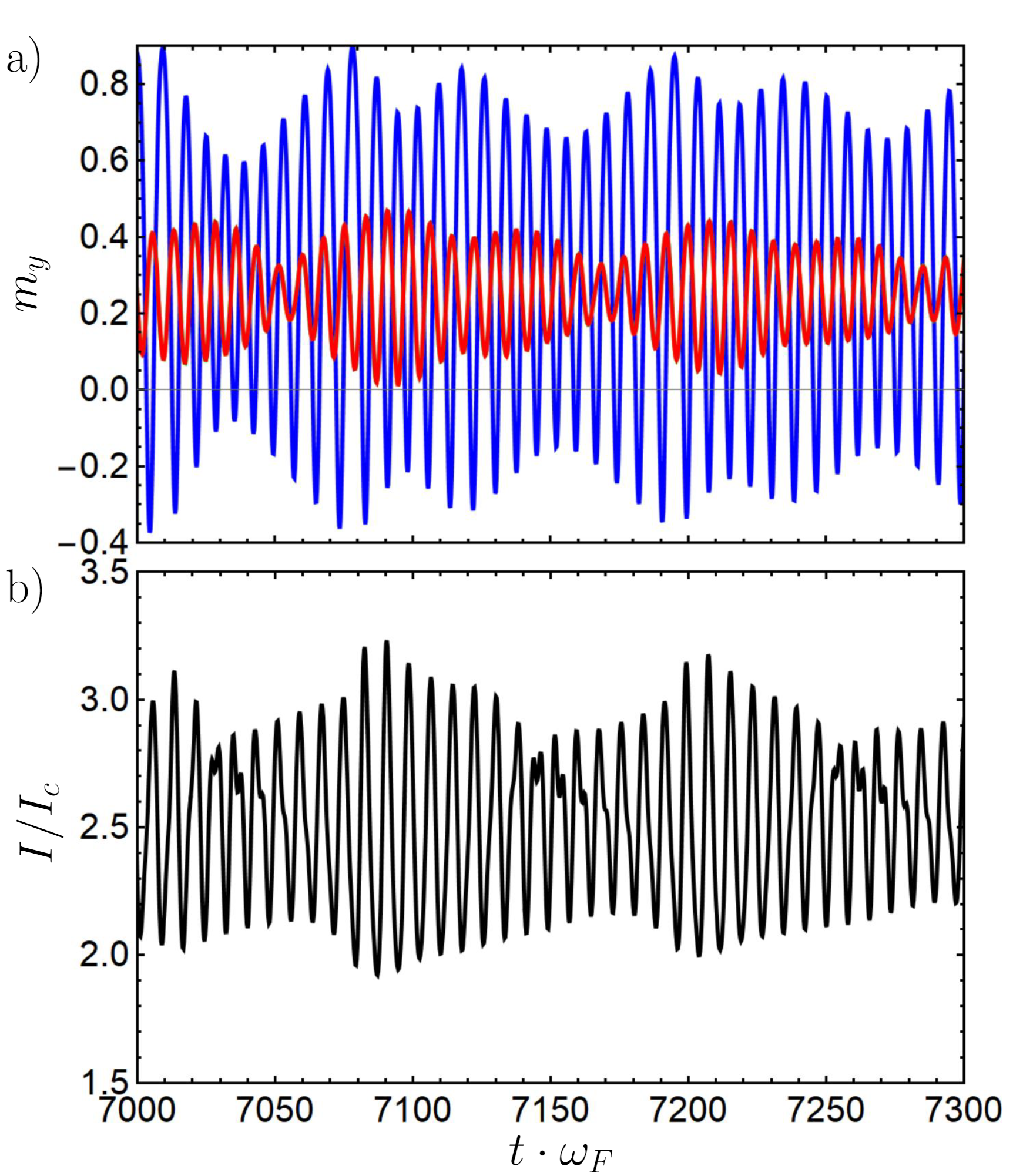}
		\caption{Regime III. (a) Dynamics of $m_{yi}$. Different colors represent different magnets. The green and red curves coincide. (b) $I_{st}(t)$. Beats are seen. The carrier frequency is very close to $\omega_J$. $eV=1.18 \hbar \omega_F$.}
  \label{fig:regimeIII}
	\end{center}
 \end{figure}

Regime III, which is realized in a narrow voltage region before regime IV is even more interesting. In this voltage region $\omega_J$ is close to the eigenfrequencies of the chain. For the case under consideration $\omega_a$ and $\omega_o$ are close to each other and close to $\omega_F$ because $r^2 E_J^0/E_M^0 \ll 1$. If this parameter were larger the voltage-driven dynamics of the system would be more trivial because the magnetic moments quickly lie along the $y$-axis due to the strong magnetoelectric action of the current-induced effective field.  According to estimates reported in the literature \cite{Konschelle2009,Bobkov2024many}, different regimes from $E_M/E_J \ll 1$ to $E_M/E_J > 1$ can be realized experimentally. $r$ also can vary widely \cite{Bobkov2024many} from $r \ll 1$ to $r \sim 10$.  Due to the proximity of the eigenfrequencies to $\omega_J$ in regime III the amplitudes of oscillations of $\bm m_i$ are large and frequent events of the magnetization reversals $\bm m_i$ along $\bm e_z \to -\bm e_z$ occur. Then the applied dc voltage is able to excite not only the acoustic but also the optic eigenmode. This is seen in Fig.~\ref{fig:regimeIII}(a), where the $y$-components of different magnetic moments $m_{yi}$ oscillate with different amplitudes, which indicates that a superposition of the acoustic and optic modes is excited. 

The current $I_{st}(t)$ in regime III, shown in Fig.~\ref{fig:regimeIII}(b) manifests beats, which are a result of oscillations with close frequencies $\omega_a$ and $\omega_o$. The same beats are seen in the dynamics of the magnetic moments. Please note that in this regime the frequency of the oscillations of $I_{st}(t)$,  $m_{yi}(t)$ and $\varphi_i(t)$ is not strictly determined by the externally forced frequency $\omega_J$, the system adjusts the frequency of oscillations by itself. This is due to the proximity to the eigenfrequencies of the magnetic system and, consequently, strong back action of the magnetic subsystem on the superconducting phase via the anomalous phase shift.  Therefore, in regime III it is possible to observe the collective magnetic eigenmodes of the coupled chain of $\varphi_0$-S/F/S JJs in the behavior of the electric current.

\section{Conclusions}
\label{conclusions}
The dc voltage-driven dynamics of a coupled chain of $\varphi_0$-S/F/S JJs is investigated. Magnetoelectric coupling between the magnetic weak links and the superconducting phase leads to an instability of the conventional phase and magnetization dynamics and to transitions to qualitatively different dynamical regimes. In particular, the collective magnetic eigenmodes of the chain can be observed as beats of the ac electric current.

\begin{acknowledgments}
The work has been supported by RSF Project No. 22-42-04408 (numerical studies of IV-characteristics and different dynamical regimes) and  by Grant from the Ministry of Science and Higher Education of the Russian Federation No. 075-15-2024-632 (analytical studies of the phase instability). 
\end{acknowledgments}

\begin{widetext}
\appendix
\section{Numerical procedure}
\label{numerical}

In this section we describe technical details of the numerical procedure. Functions to be calculated are $\bm m_i(t),\varphi_i(t)$. In total we have $3N$ dynamical variables for magnetization vectors and $N$ variables for phase differences with $N$ constraints $|\bm m_i|=1$ for magnetization vectors and one constraint $\sum \limits_{i=1}^N \varphi_i = \psi_N = 2eVt/\hbar$. Equations (3) and (4) of the main text can be transformed to the following form (the standard form of the initial value problem) to facilitate numerical calculations:

\begin{eqnarray}
    \dot {\bm m}_i=-\frac{\gamma}{1+\alpha^2}[\bm m_i\times \bm H^i]- 
    \frac{\alpha\gamma}{1+\alpha^2}[\bm m_i\times [\bm m_i \times \bm H^i]]
    \label{RK1}
\end{eqnarray}

\begin{eqnarray}
    \dot \varphi_i=\frac{2eR_N}{\hbar}\left( I(t) -I_{c,i} \sin(\varphi_i-r_i m_{iy})\right)+r_i \dot { m}_{iy}~~
    \label{RK2}
\end{eqnarray}

where

\begin{eqnarray}
    \bm H^i=\bm H^i_M+\frac{r_i\hbar}{2e M V_F} I(t) \bm e_y .
    \label{eq:A3}
\end{eqnarray}
Then we need $I(t)$ in terms of $\bm m_i(t)$ and $\varphi_i(t)$. It can be obtained by summing Eq.~(\ref{RK2}) over all $i$, which results in
\begin{eqnarray}
    \frac{V}{R_N}=N I(t)-\sum\limits_i \left[ I_{c,i}\sin(\varphi_i-r_i m_{iy}) -\frac{\hbar}{2eR_N}r_i \dot { m}_{iy} \right]
\end{eqnarray}
Making use of Eq.~(\ref{RK1}) $\dot { m}_{iy}$ can be expressed in terms of $\bm m_i$ and $I(t)$. The final expression for $I(t)$ takes the form:

\begin{eqnarray}
I(t)=\frac{\frac{V}{R_N}+\sum\limits_i \left[ I_{c,i}\sin(\varphi_i-r_i m_{iy})+\frac{\hbar r_i\gamma}{2eR_N(1+\alpha^2)}\left(\bm e_y( [\bm m_i\times \bm H^i_M]+\alpha [\bm m_i \times [\bm m_i\times \bm H^i_M]] )\right)\right ]}{N-\sum\limits_i \frac{\hbar^2 \gamma r_i^2 }{4e^2 R_N M V_F (1+\alpha^2)}\left( \bm e_y ([\bm m_i\times \bm e_y]+\alpha [\bm m_i\times [\bm m_i\times \bm e_y] ]
) \right)}
\end{eqnarray}

\begin{figure}[tb]
	\begin{center}		\includegraphics[width=75mm]{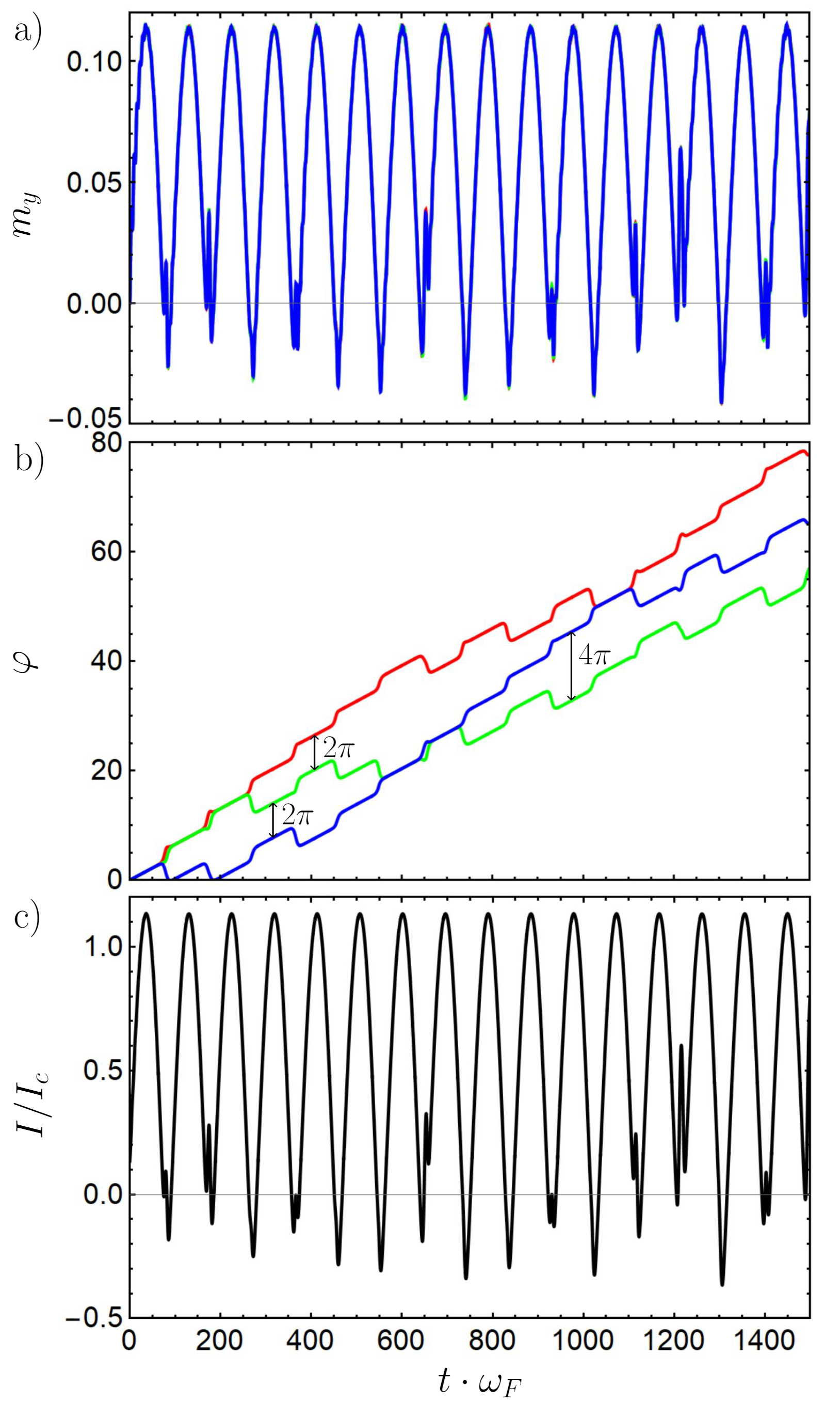}
		\caption{Regime I. (a) $m_y(t)$ [the same at all three JJs of the chain with $N=3$]. (b) Dynamics of the phase differences $\varphi_i$ at all three JJs. (c) Oscillations of $I(t)$ with frequency $\omega_J$. $eV=0.067 \hbar \omega_F$}
  \label{fig:phase_slips}
	\end{center}
 \end{figure}

Equations (\ref{RK1}), (\ref{RK2}) are solved by the explicit second order Runge-Kutta method. The time step is $\delta t=0.002 \omega_F^{-1}$, where $\omega_F = \gamma (K/M)$ is the ferromagnetic resonance frequency. Time domain $10^{4} \omega_F^{-1}$ was used to be sure that all transient processes are ended. 

The initial conditions are $\bm m_i=\bm e_z$, $\varphi_i=0$ which corresponds to the equilibrium state before switching on the voltage. The voltage is switched on abruptly, but it has been verified that the result (except for the transition period at times $t\sim \omega_F^{-1}$) does not depend on the method of switching on the voltage.

To add small noise to the calculation, the following term was added to the effective field:

\begin{eqnarray}
    \delta \bm H^i=\frac{P_{\rm noise}}{\sqrt{dt}} \bm f_{\rm rnd}
\end{eqnarray}
where $\bm f_{\rm rnd}$ is a random Gaussian variable with unit variance. The amplitude of noise $P_{\rm noise}=0.0002 K/(M\sqrt{\omega_F})$ was used in our calculations.

\section{Low-voltage phase-slips regime}

\label{low-voltage}

\begin{figure}[tb]
	\begin{center}		\includegraphics[width=115mm]{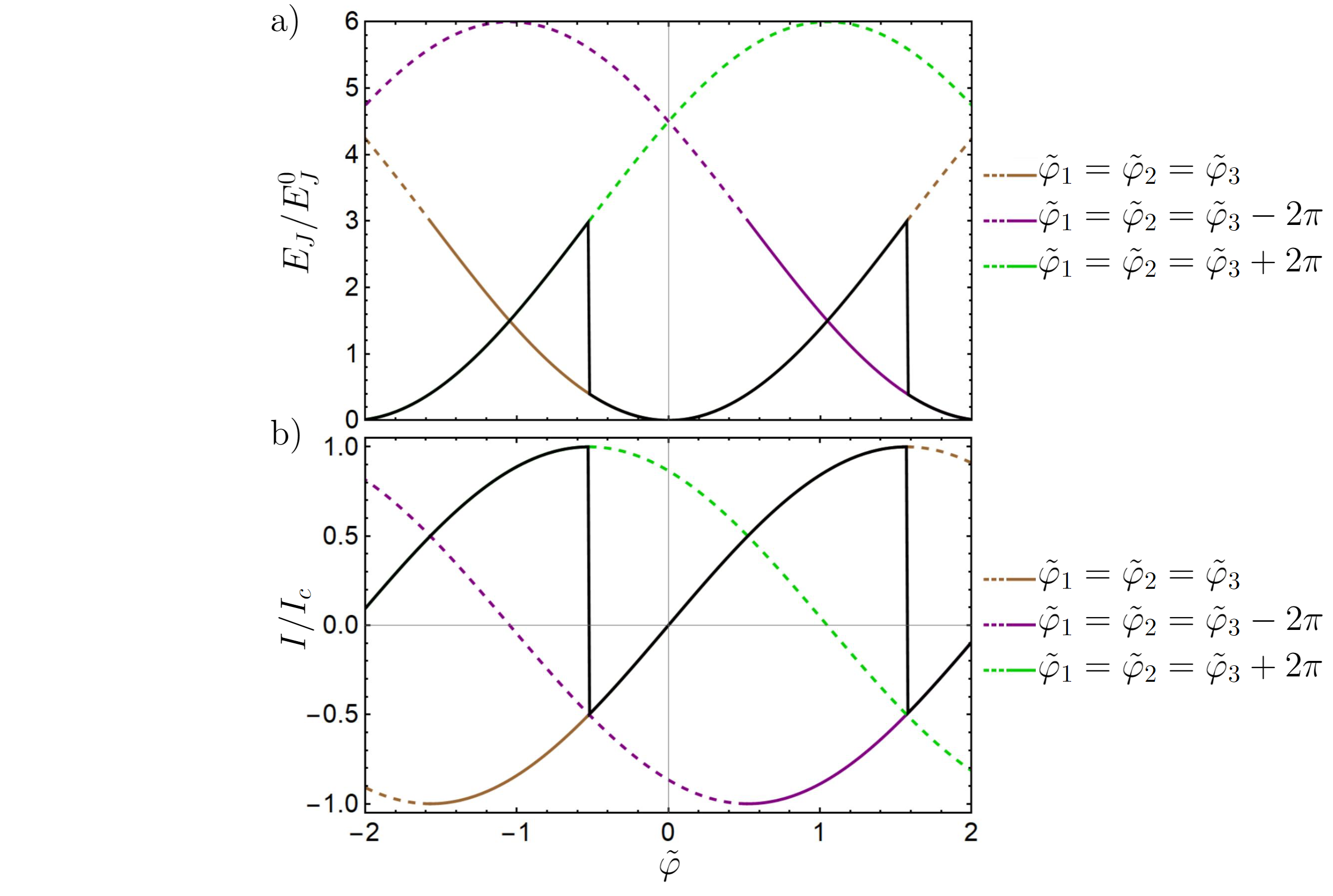}
		\caption{(a) Josephson energy $E_J (\tilde \varphi)$ of the coupled chain of JJs. Stable parts of the energy branches are shown by solid lines, and unstable with respect to phase slips parts are shown by dashed lines. The abrupt jumps of the black line correspond to phase slips. (b) Josephson current $I(\tilde \varphi)$. The black lines in the both panels illustrates the dynamical variation of the chain Josephson energy and the Josephson current in the adiabatic approximation.}
  \label{fig:phase_slips2}
	\end{center}
 \end{figure}
 \begin{figure}[tb]
	\begin{center}		\includegraphics[width=68mm]{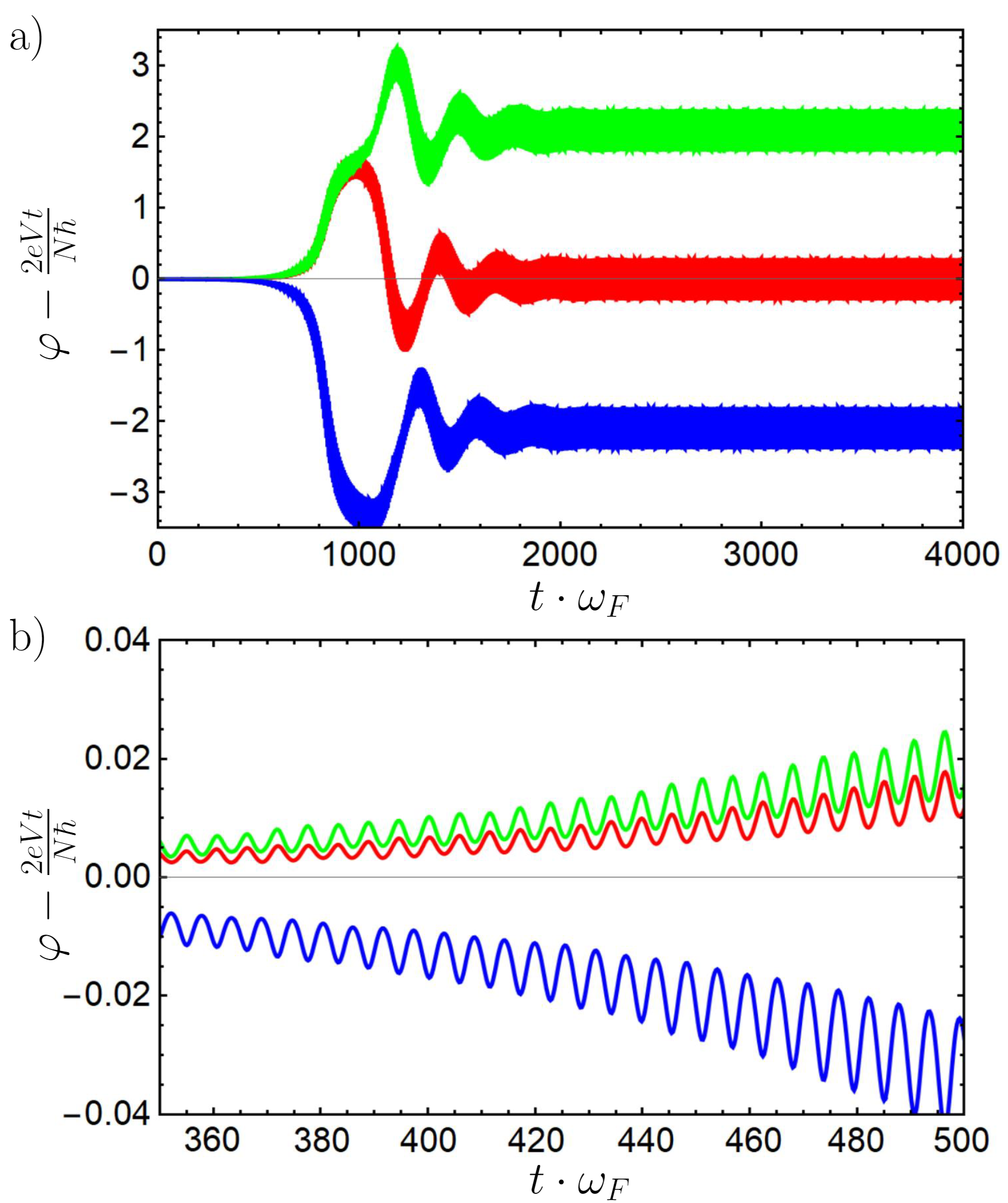}
		\caption{Regime IV. (a) Dynamics of the phase differences $\varphi_i$ at all three JJs of the chain with $N=3$. (b) The same on a larger scale. $eV=1.67 \hbar \omega_F$.}
  \label{fig:instability}
	\end{center}
 \end{figure}
As  was discussed in the main text, in low-voltage regime I the dynamics of the phase differences $\varphi_i (t)$ is determined by phase slips. Here by an example of a chain consisting of $N=3$ coupled $\varphi_0$-S/F/S JJs we provide more detailed numerical results on the dynamics of the phase and the electric current in this regime. The time dependence of the phase differences $\varphi_i(t)$ at all three JJs is shown in Fig.~\ref{fig:phase_slips}. The regular linear growth is accompanied by the phase slips $\varphi_i - \varphi_j = 2\pi n$, where $n$ is an integer number. This condition provides the current conservation $I_i(t) = I_j(t)$. Also the condition $\sum \limits_i \varphi_i = 2eVt/\hbar$ is fulfilled. 

The physical reason for the classical phase slips is well-known [see, for example, Ref.~\onlinecite{Pop2010}] and explained in Fig.~\ref{fig:phase_slips2} by considering the adiabatic limit corresponding to small $V$. The Josephson energy of the system takes the form:
\begin{eqnarray}
E_J=\sum \limits_i E_J^0 (1-\cos \tilde \varphi_i),
\label{suppl:energy}
\end{eqnarray}
where $\tilde \varphi_i = \tilde \varphi + \delta \varphi_i $. $\sum \limits_i \delta \varphi_i = 0$ and $\tilde \varphi = 2eV t/N \hbar - \varphi_{0}$ is the externally applied phase equally distributed between all JJs shifted by the anomalous phase shift $\varphi_0$, which is the same for all JJs in the adiabatic approximation. The energy is shown in Fig.~\ref{fig:phase_slips2}(a). It is a multi-valued function of $\tilde \varphi$. The unstable with respect to phase-slips parts of the energy branches are shown by dashed lines. The higher energetically unfavorable parts of the branches are partially metastable. As a result the jumps between the branches occur later their crossing points $\omega_J t = (2k+1)\pi/N$, where $k=0,\pm 1,...$. Due to this fact the dependence $I(\tilde \varphi)$ is asymmetric with respect to the horizontal axis, see Fig.~\ref{fig:phase_slips2}(b) ($\dot {\tilde \varphi}$ is neglected in this analysis). This is the reason for the excess current with respect to  Ohm's law, which is seen in the IV-characteristics, presented in Fig.~2 of the main text.

In order to obtain numerically the results presented in Fig.~\ref{fig:phase_slips} we assumed small noise of the magnetic moments $
\bm m_i$. But we would like to stress that the discussed results are not related to the magnetoelectric coupling between the phase and the magnetization and can be realized even in coupled chains of conventional JJs with $\varphi_0 \equiv 0$ if we assume small fluctuations of the phases. In general, due to the finite value of the internal resistance of the generator some small initial part of regime I at $V \sim I_c r_g$ may be difficult to achieve experimentally.

\section{Phase instability}

\label{instability}

\begin{figure}[tb]
	\begin{center}		\includegraphics[width=165mm]{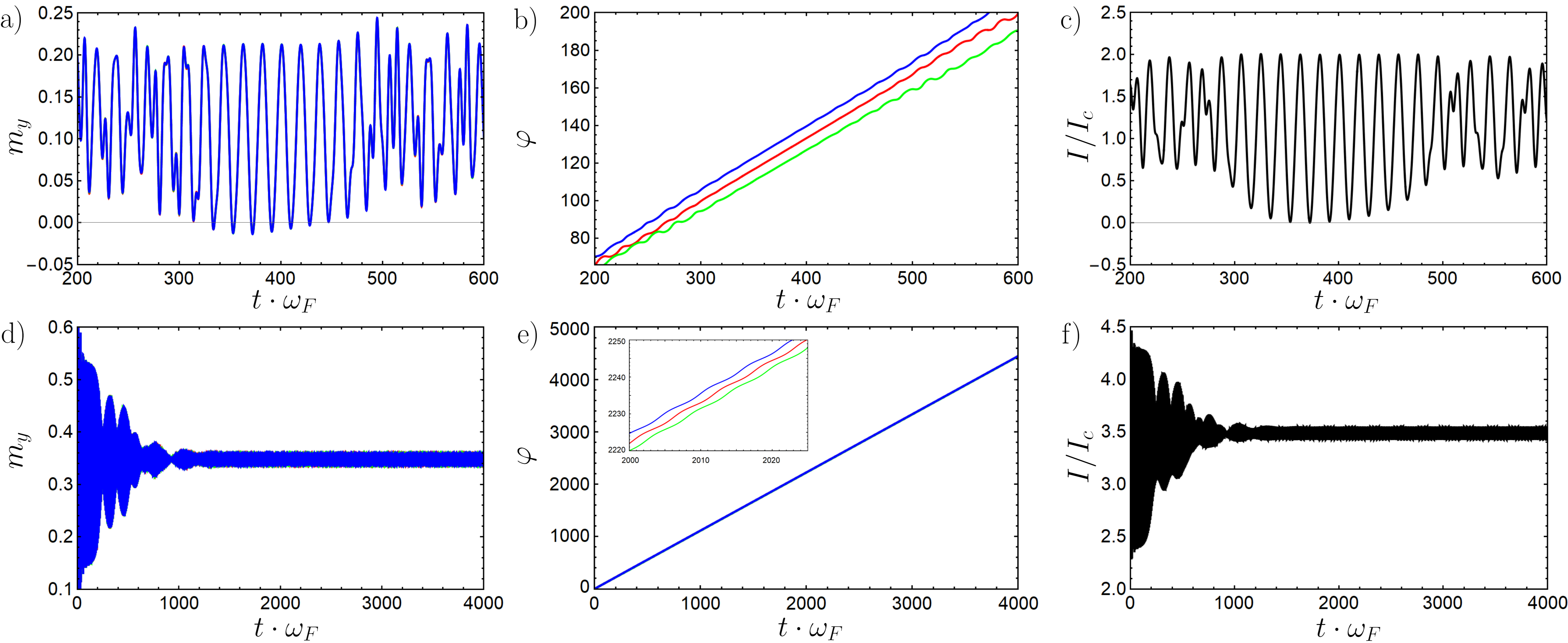}
		\caption{Dynamics of $\varphi_0$-S/F/S JJs chain with $N=3$, where the critical currents of individual JJs are not the same. $\delta I_c/I_c=2\%$. Upper row: regime II. (a) $m_{yi}(t)$ [the same for all JJs]; (b) $\varphi_i(t)$  and (c) $I_{st}(t)$ at $eV=0.5\hbar \omega_F$. Only the stationary state is shown. Bottom row: regime IV. (d) $m_{yi}(t)$ [the same for all JJs]; (e) $\varphi_i(t)$, which are different for all magnets, which is shown on a larger scale in the inset. (f) $I(t)$ at $eV=1.67 \hbar \omega_F$. The full time range is shown including the initial and transitional regimes.}
  \label{fig:different_1}
	\end{center}
 \end{figure}

 \begin{figure}[tb]
	\begin{center}		\includegraphics[width=165mm]{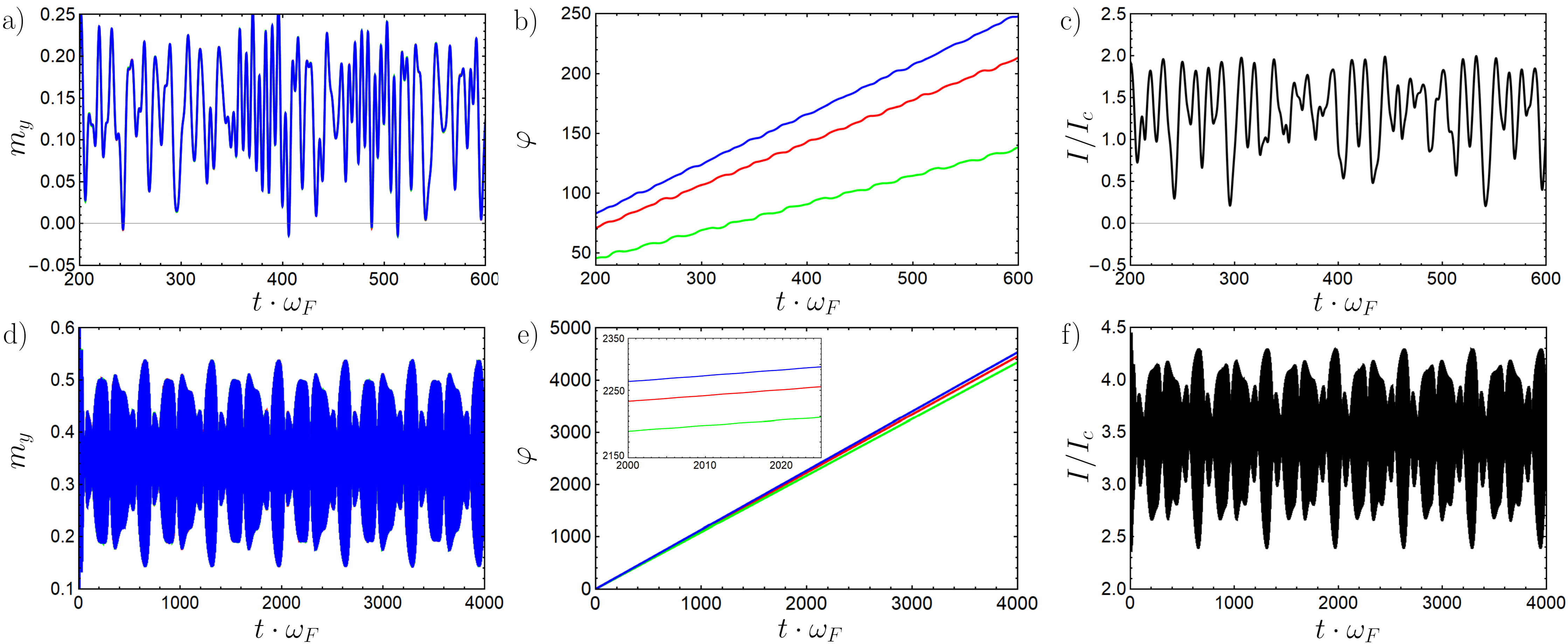}
		\caption{The same as in Fig.~\ref{fig:different_1}, but for much larger variations of the critical currents $\delta I_c/I_c=35\%$.}
  \label{fig:different_2}
	\end{center}
 \end{figure}

Here we discuss in detail the process of increasing phase instability and transition of the system to the new regime IV discussed in the main text. We apply voltage $V \neq 0$ at $t=0$.  The dynamics of the phase differences at all JJs is shown in Fig.~\ref{fig:instability} for $N=3$.  At small times all the phase differences are the same and correspond to the linear growth $\varphi_i = \omega_J t$ like in the regular regime II. However, the considered voltage exceeds the eigenfrequencies $\omega_{a,o}$ of the chain. Then we have $\kappa<0$ in Eq.~(8) of the main text and this regular growth of the phase difference becomes unstable. The process of growing instability is seen in Fig.~\ref{fig:instability}(a) and in Fig.~\ref{fig:instability}(b) on a larger scale. Its typical time exceeds all the characteristic time scales of the system and depends not only on the system parameters, but also on amplitude of the initial fluctuations of $\bm m_i$. After the system comes to the stationary state, the quantities $\varphi_i - \omega_J t$ oscillate with the same frequency $\omega_J$, but they have different time-averaged values and there is a constant phase shift between their oscillations. We have numerically checked that the final dynamical state of the system is not unique. The phase differences always oscillate with frequency $\omega_J$, but their time-averaged values and the phase shifts between them depend on the particular realization. Surely, all possible realizations presuppose the fulfillment of the condition $\sum \limits \varphi_i (t) = 2eVt/\hbar$. 

\section{Influence of variations of individual JJs parameters}
\label{variations}

\begin{figure}[tb]
	\begin{center}		\includegraphics[width=75mm]{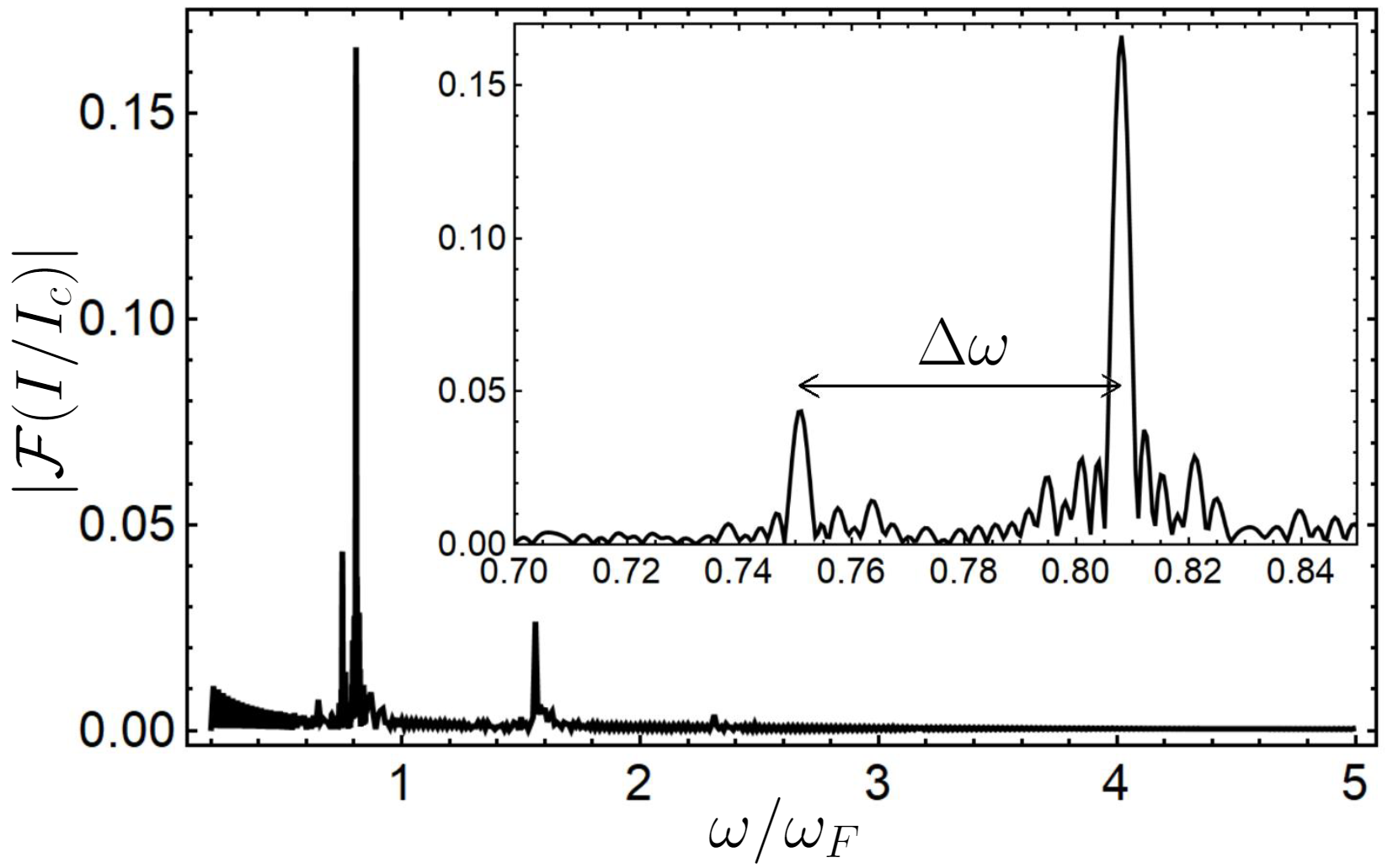}
		\caption{Fourier transform of $I(t)$ presented in Fig.~5 of the main text. Inset: the region of the split main peak on a larger scale. $eV=1.18\hbar\omega_F$.}
  \label{fig:FFT}
	\end{center}
 \end{figure}

In this section we present results supporting our statement that variations of parameters of individual JJs do not prevent the system from transitioning to anomalous dynamic regimes. We assume that in the chain consisting of $N=3$ junctions individual JJs have different critical currents $I_c$, $I_c \pm \delta I_c$. Figs.~\ref{fig:different_1} and \ref{fig:different_2} demonstrate the results for the voltage-driven dynamics  in the normal regime II (upper row) and the anomalous regime IV (bottom row). In Fig.~~\ref{fig:different_1} the case of small critical current variations $\delta I_c/I_c=2\%$ is shown, while Fig.~~\ref{fig:different_2} corresponds to the case of much larger $\delta I_c/I_c=35\%$.  Due to the different values of the critical currents the normal regime in  both figures is characterized by different phase differences at all three JJs and, consequently, different inclines of $\varphi_i(t)$. In Fig.~\ref{fig:different_2}(b) it is seen more clearly than in Fig.~\ref{fig:different_1}(b) owing to the stronger variation of the critical currents. Because of different values $\dot \varphi_i(t)$ the Josephson frequencies of the individual JJs are not the same. If this difference is small, it results in beats of the oscillations of $m_{yi}(t)$ and $I(t)$, see Figs.~\ref{fig:different_2}(a) and (c). For larger differencies of the Josephson frequencies of individual JJs the behavior  of $m_{yi}(t)$ and $I(t)$ is more complicated, see Figs.~\ref{fig:different_2}(d) and (f).

At the same time the anomalous regime IV is characterized by much smaller differences of the averaged incline of $\varphi_i(t)$ and smaller amplitudes of the $I(t)$ oscillations. For not very large variations of the critical currents the amplitudes of the oscillations of the magnetic moments are also suppressed in this regime. In the anomalous regime III observation of the beats corresponding to the difference of the eigenfrequencies of the chain collective modes $\omega_a-\omega_o$ is complicated by the simultaneous presence of the beats associated with the difference of the Josephson frequencies of the individual JJs.

\section{Fourier analysis of current beats}

Eqs. (5),(6) of the main text give us the difference between optic and acoustic modes

\begin{eqnarray}
    \Delta \omega =\omega_a-\omega_o=\omega_F \left(\sqrt{1+\frac{2r^2E_J^0}{E_M^0}}-1\right)
    \label{delta_omega}
\end{eqnarray}
For the parameters used in Fig.~5 of the main text $\Delta \omega\approx 0.06\omega_F$. In Fig.\ref{fig:FFT} the Fourier transform of beats of current is shown. The main peak is doubled (also peaks corresponding to multiple frequencies can be seen). The difference between the two frequencies that make up the beats is $0.057\omega_F$, which agrees with Eq.~\ref{delta_omega} very well. 

It is important to note that since the eigenfrequencies depend on $\psi_N$, the frequencies $\omega_a,\omega_o$ by themselves  are not determined by Eqs.~(5) and (6) of the main text,  as it is indicated there. Nevertheless, the expression for their difference Eq.~\ref{delta_omega} remains valid.

\end{widetext}

\bibliography{voltage}

%apsrev4-2.bst 2019-01-14 (MD) hand-edited version of apsrev4-1.bst
%Control: key (0)
%Control: author (8) initials jnrlst
%Control: editor formatted (1) identically to author
%Control: production of article title (0) allowed
%Control: page (0) single
%Control: year (1) truncated
%Control: production of eprint (0) enabled
\begin{thebibliography}{54}%
\makeatletter
\providecommand \@ifxundefined [1]{%
 \@ifx{#1\undefined}
}%
\providecommand \@ifnum [1]{%
 \ifnum #1\expandafter \@firstoftwo
 \else \expandafter \@secondoftwo
 \fi
}%
\providecommand \@ifx [1]{%
 \ifx #1\expandafter \@firstoftwo
 \else \expandafter \@secondoftwo
 \fi
}%
\providecommand \natexlab [1]{#1}%
\providecommand \enquote  [1]{``#1''}%
\providecommand \bibnamefont  [1]{#1}%
\providecommand \bibfnamefont [1]{#1}%
\providecommand \citenamefont [1]{#1}%
\providecommand \href@noop [0]{\@secondoftwo}%
\providecommand \href [0]{\begingroup \@sanitize@url \@href}%
\providecommand \@href[1]{\@@startlink{#1}\@@href}%
\providecommand \@@href[1]{\endgroup#1\@@endlink}%
\providecommand \@sanitize@url [0]{\catcode `\\12\catcode `\$12\catcode `\&12\catcode `\#12\catcode `\^12\catcode `\_12\catcode `\%12\relax}%
\providecommand \@@startlink[1]{}%
\providecommand \@@endlink[0]{}%
\providecommand \url  [0]{\begingroup\@sanitize@url \@url }%
\providecommand \@url [1]{\endgroup\@href {#1}{\urlprefix }}%
\providecommand \urlprefix  [0]{URL }%
\providecommand \Eprint [0]{\href }%
\providecommand \doibase [0]{https://doi.org/}%
\providecommand \selectlanguage [0]{\@gobble}%
\providecommand \bibinfo  [0]{\@secondoftwo}%
\providecommand \bibfield  [0]{\@secondoftwo}%
\providecommand \translation [1]{[#1]}%
\providecommand \BibitemOpen [0]{}%
\providecommand \bibitemStop [0]{}%
\providecommand \bibitemNoStop [0]{.\EOS\space}%
\providecommand \EOS [0]{\spacefactor3000\relax}%
\providecommand \BibitemShut  [1]{\csname bibitem#1\endcsname}%
\let\auto@bib@innerbib\@empty
%</preamble>
\bibitem [{\citenamefont {Golubov}\ \emph {et~al.}(2004)\citenamefont {Golubov}, \citenamefont {Kupriyanov},\ and\ \citenamefont {Il'ichev}}]{Golubov_review}%
  \BibitemOpen
  \bibfield  {author} {\bibinfo {author} {\bibfnamefont {A.~A.}\ \bibnamefont {Golubov}}, \bibinfo {author} {\bibfnamefont {M.~Y.}\ \bibnamefont {Kupriyanov}},\ and\ \bibinfo {author} {\bibfnamefont {E.}~\bibnamefont {Il'ichev}},\ }\bibfield  {title} {\bibinfo {title} {The current-phase relation in josephson junctions},\ }\href {https://doi.org/10.1103/RevModPhys.76.411} {\bibfield  {journal} {\bibinfo  {journal} {Rev. Mod. Phys.}\ }\textbf {\bibinfo {volume} {76}},\ \bibinfo {pages} {411} (\bibinfo {year} {2004})}\BibitemShut {NoStop}%
\bibitem [{\citenamefont {Buzdin}\ \emph {et~al.}(1982)\citenamefont {Buzdin}, \citenamefont {Bulaevskii},\ and\ \citenamefont {V.}}]{Buzdin1982}%
  \BibitemOpen
  \bibfield  {author} {\bibinfo {author} {\bibfnamefont {A.~I.}\ \bibnamefont {Buzdin}}, \bibinfo {author} {\bibfnamefont {L.~N.}\ \bibnamefont {Bulaevskii}},\ and\ \bibinfo {author} {\bibfnamefont {P.~S.}\ \bibnamefont {V.}},\ }\bibfield  {title} {\bibinfo {title} {Critical-current oscillations as a function of the exchange field and thickness of the ferromagnetic metal (f) in an s-f-s josephson junction},\ }\href {http://jetpletters.ru/ps/1314/article_19853.shtml} {\bibfield  {journal} {\bibinfo  {journal} {JETP Lett.}\ }\textbf {\bibinfo {volume} {35}},\ \bibinfo {pages} {178} (\bibinfo {year} {1982})}\BibitemShut {NoStop}%
\bibitem [{\citenamefont {Buzdin}(2005)}]{Buzdin2005}%
  \BibitemOpen
  \bibfield  {author} {\bibinfo {author} {\bibfnamefont {A.~I.}\ \bibnamefont {Buzdin}},\ }\bibfield  {title} {\bibinfo {title} {Proximity effects in superconductor-ferromagnet heterostructures},\ }\href {https://doi.org/10.1103/RevModPhys.77.935} {\bibfield  {journal} {\bibinfo  {journal} {Rev. Mod. Phys.}\ }\textbf {\bibinfo {volume} {77}},\ \bibinfo {pages} {935} (\bibinfo {year} {2005})}\BibitemShut {NoStop}%
\bibitem [{\citenamefont {Ryazanov}\ \emph {et~al.}(2001)\citenamefont {Ryazanov}, \citenamefont {Oboznov}, \citenamefont {Rusanov}, \citenamefont {Veretennikov}, \citenamefont {Golubov},\ and\ \citenamefont {Aarts}}]{Ryazanov2001}%
  \BibitemOpen
  \bibfield  {author} {\bibinfo {author} {\bibfnamefont {V.~V.}\ \bibnamefont {Ryazanov}}, \bibinfo {author} {\bibfnamefont {V.~A.}\ \bibnamefont {Oboznov}}, \bibinfo {author} {\bibfnamefont {A.~Y.}\ \bibnamefont {Rusanov}}, \bibinfo {author} {\bibfnamefont {A.~V.}\ \bibnamefont {Veretennikov}}, \bibinfo {author} {\bibfnamefont {A.~A.}\ \bibnamefont {Golubov}},\ and\ \bibinfo {author} {\bibfnamefont {J.}~\bibnamefont {Aarts}},\ }\bibfield  {title} {\bibinfo {title} {Coupling of two superconductors through a ferromagnet: Evidence for a $\ensuremath{\pi}$ junction},\ }\href {https://doi.org/10.1103/PhysRevLett.86.2427} {\bibfield  {journal} {\bibinfo  {journal} {Phys. Rev. Lett.}\ }\textbf {\bibinfo {volume} {86}},\ \bibinfo {pages} {2427} (\bibinfo {year} {2001})}\BibitemShut {NoStop}%
\bibitem [{\citenamefont {Kontos}\ \emph {et~al.}(2002)\citenamefont {Kontos}, \citenamefont {Aprili}, \citenamefont {Lesueur}, \citenamefont {Gen\^et}, \citenamefont {Stephanidis},\ and\ \citenamefont {Boursier}}]{Kontos2002}%
  \BibitemOpen
  \bibfield  {author} {\bibinfo {author} {\bibfnamefont {T.}~\bibnamefont {Kontos}}, \bibinfo {author} {\bibfnamefont {M.}~\bibnamefont {Aprili}}, \bibinfo {author} {\bibfnamefont {J.}~\bibnamefont {Lesueur}}, \bibinfo {author} {\bibfnamefont {F.}~\bibnamefont {Gen\^et}}, \bibinfo {author} {\bibfnamefont {B.}~\bibnamefont {Stephanidis}},\ and\ \bibinfo {author} {\bibfnamefont {R.}~\bibnamefont {Boursier}},\ }\bibfield  {title} {\bibinfo {title} {Josephson junction through a thin ferromagnetic layer: Negative coupling},\ }\href {https://doi.org/10.1103/PhysRevLett.89.137007} {\bibfield  {journal} {\bibinfo  {journal} {Phys. Rev. Lett.}\ }\textbf {\bibinfo {volume} {89}},\ \bibinfo {pages} {137007} (\bibinfo {year} {2002})}\BibitemShut {NoStop}%
\bibitem [{\citenamefont {Robinson}\ \emph {et~al.}(2006)\citenamefont {Robinson}, \citenamefont {Piano}, \citenamefont {Burnell}, \citenamefont {Bell},\ and\ \citenamefont {Blamire}}]{robinson2006critical}%
  \BibitemOpen
  \bibfield  {author} {\bibinfo {author} {\bibfnamefont {J.}~\bibnamefont {Robinson}}, \bibinfo {author} {\bibfnamefont {S.}~\bibnamefont {Piano}}, \bibinfo {author} {\bibfnamefont {G.}~\bibnamefont {Burnell}}, \bibinfo {author} {\bibfnamefont {C.}~\bibnamefont {Bell}},\ and\ \bibinfo {author} {\bibfnamefont {M.}~\bibnamefont {Blamire}},\ }\bibfield  {title} {\bibinfo {title} {Critical current oscillations in strong ferromagnetic $\pi$ junctions},\ }\href {https://journals.aps.org/prl/abstract/10.1103/PhysRevLett.97.177003} {\bibfield  {journal} {\bibinfo  {journal} {Physical review letters}\ }\textbf {\bibinfo {volume} {97}},\ \bibinfo {pages} {177003} (\bibinfo {year} {2006})}\BibitemShut {NoStop}%
\bibitem [{\citenamefont {Baselmans}\ \emph {et~al.}(1999)\citenamefont {Baselmans}, \citenamefont {Morpurgo}, \citenamefont {van Wees},\ and\ \citenamefont {Klapwijk}}]{Baselmans1999}%
  \BibitemOpen
  \bibfield  {author} {\bibinfo {author} {\bibfnamefont {J.~J.~A.}\ \bibnamefont {Baselmans}}, \bibinfo {author} {\bibfnamefont {A.~F.}\ \bibnamefont {Morpurgo}}, \bibinfo {author} {\bibfnamefont {B.~J.}\ \bibnamefont {van Wees}},\ and\ \bibinfo {author} {\bibfnamefont {T.~M.}\ \bibnamefont {Klapwijk}},\ }\bibfield  {title} {\bibinfo {title} {Reversing the direction of the supercurrent in a controllable josephson junction},\ }\href {https://doi.org/10.1038/16204} {\bibfield  {journal} {\bibinfo  {journal} {Nature}\ }\textbf {\bibinfo {volume} {397}},\ \bibinfo {pages} {43} (\bibinfo {year} {1999})}\BibitemShut {NoStop}%
\bibitem [{\citenamefont {Golikova}\ \emph {et~al.}(2021)\citenamefont {Golikova}, \citenamefont {Wolf}, \citenamefont {Beckmann}, \citenamefont {Penzyakov}, \citenamefont {Batov}, \citenamefont {Bobkova}, \citenamefont {Bobkov},\ and\ \citenamefont {Ryazanov}}]{golikova2021controllable}%
  \BibitemOpen
  \bibfield  {author} {\bibinfo {author} {\bibfnamefont {T.~E.}\ \bibnamefont {Golikova}}, \bibinfo {author} {\bibfnamefont {M.~J.}\ \bibnamefont {Wolf}}, \bibinfo {author} {\bibfnamefont {D.}~\bibnamefont {Beckmann}}, \bibinfo {author} {\bibfnamefont {G.~A.}\ \bibnamefont {Penzyakov}}, \bibinfo {author} {\bibfnamefont {I.~E.}\ \bibnamefont {Batov}}, \bibinfo {author} {\bibfnamefont {I.}~\bibnamefont {Bobkova}}, \bibinfo {author} {\bibfnamefont {A.~M.}\ \bibnamefont {Bobkov}},\ and\ \bibinfo {author} {\bibfnamefont {V.~V.}\ \bibnamefont {Ryazanov}},\ }\bibfield  {title} {\bibinfo {title} {Controllable supercurrent in mesoscopic superconductor-normal metal-ferromagnetcrosslike josephson structures},\ }\href {https://iopscience.iop.org/article/10.1088/1361-6668/abfd0d/meta} {\bibfield  {journal} {\bibinfo  {journal} {Superconductor Science and Technology}\ } (\bibinfo {year} {2021})}\BibitemShut {NoStop}%
\bibitem [{\citenamefont {Schulz}\ \emph {et~al.}(2000)\citenamefont {Schulz}, \citenamefont {Chesca}, \citenamefont {G{\"o}tz}, \citenamefont {Schneider}, \citenamefont {Schmehl}, \citenamefont {Bielefeldt}, \citenamefont {Hilgenkamp}, \citenamefont {Mannhart},\ and\ \citenamefont {Tsuei}}]{schulz2000design}%
  \BibitemOpen
  \bibfield  {author} {\bibinfo {author} {\bibfnamefont {R.~R.}\ \bibnamefont {Schulz}}, \bibinfo {author} {\bibfnamefont {B.}~\bibnamefont {Chesca}}, \bibinfo {author} {\bibfnamefont {B.}~\bibnamefont {G{\"o}tz}}, \bibinfo {author} {\bibfnamefont {C.~W.}\ \bibnamefont {Schneider}}, \bibinfo {author} {\bibfnamefont {A.}~\bibnamefont {Schmehl}}, \bibinfo {author} {\bibfnamefont {H.}~\bibnamefont {Bielefeldt}}, \bibinfo {author} {\bibfnamefont {H.}~\bibnamefont {Hilgenkamp}}, \bibinfo {author} {\bibfnamefont {J.}~\bibnamefont {Mannhart}},\ and\ \bibinfo {author} {\bibfnamefont {C.}~\bibnamefont {Tsuei}},\ }\bibfield  {title} {\bibinfo {title} {Design and realization of an all d-wave dc $\pi$-superconducting quantum interference device},\ }\href {https://aip.scitation.org/doi/abs/10.1063/1.125627} {\bibfield  {journal} {\bibinfo  {journal} {Applied Physics Letters}\ }\textbf {\bibinfo {volume} {76}},\ \bibinfo {pages} {912} (\bibinfo {year} {2000})}\BibitemShut {NoStop}%
\bibitem [{\citenamefont {J{\o}rgensen}\ \emph {et~al.}(2007)\citenamefont {J{\o}rgensen}, \citenamefont {Novotn{\`y}}, \citenamefont {Grove-Rasmussen}, \citenamefont {Flensberg},\ and\ \citenamefont {Lindelof}}]{jorgensen2007critical}%
  \BibitemOpen
  \bibfield  {author} {\bibinfo {author} {\bibfnamefont {H.~I.}\ \bibnamefont {J{\o}rgensen}}, \bibinfo {author} {\bibfnamefont {T.}~\bibnamefont {Novotn{\`y}}}, \bibinfo {author} {\bibfnamefont {K.}~\bibnamefont {Grove-Rasmussen}}, \bibinfo {author} {\bibfnamefont {K.}~\bibnamefont {Flensberg}},\ and\ \bibinfo {author} {\bibfnamefont {P.}~\bibnamefont {Lindelof}},\ }\bibfield  {title} {\bibinfo {title} {Critical current 0- $\pi$ transition in designed josephson quantum dot junctions},\ }\href {https://pubs.acs.org/doi/10.1021/nl071152w} {\bibfield  {journal} {\bibinfo  {journal} {Nano letters}\ }\textbf {\bibinfo {volume} {7}},\ \bibinfo {pages} {2441} (\bibinfo {year} {2007})}\BibitemShut {NoStop}%
\bibitem [{\citenamefont {van Dam}\ \emph {et~al.}(2006)\citenamefont {van Dam}, \citenamefont {Nazarov}, \citenamefont {Bakkers}, \citenamefont {De~Franceschi},\ and\ \citenamefont {Kouwenhoven}}]{vanDam2006}%
  \BibitemOpen
  \bibfield  {author} {\bibinfo {author} {\bibfnamefont {J.~A.}\ \bibnamefont {van Dam}}, \bibinfo {author} {\bibfnamefont {Y.~V.}\ \bibnamefont {Nazarov}}, \bibinfo {author} {\bibfnamefont {E.~P. A.~M.}\ \bibnamefont {Bakkers}}, \bibinfo {author} {\bibfnamefont {S.}~\bibnamefont {De~Franceschi}},\ and\ \bibinfo {author} {\bibfnamefont {L.~P.}\ \bibnamefont {Kouwenhoven}},\ }\bibfield  {title} {\bibinfo {title} {Supercurrent reversal in quantum dots},\ }\href {https://doi.org/10.1038/nature05018} {\bibfield  {journal} {\bibinfo  {journal} {Nature}\ }\textbf {\bibinfo {volume} {442}},\ \bibinfo {pages} {667} (\bibinfo {year} {2006})}\BibitemShut {NoStop}%
\bibitem [{\citenamefont {Ke}\ \emph {et~al.}(2019)\citenamefont {Ke}, \citenamefont {Moehle}, \citenamefont {de~Vries}, \citenamefont {Thomas}, \citenamefont {Metti}, \citenamefont {Guinn}, \citenamefont {Kallaher}, \citenamefont {Lodari}, \citenamefont {Scappucci}, \citenamefont {Wang} \emph {et~al.}}]{ke2019ballistic}%
  \BibitemOpen
  \bibfield  {author} {\bibinfo {author} {\bibfnamefont {C.~T.}\ \bibnamefont {Ke}}, \bibinfo {author} {\bibfnamefont {C.~M.}\ \bibnamefont {Moehle}}, \bibinfo {author} {\bibfnamefont {F.~K.}\ \bibnamefont {de~Vries}}, \bibinfo {author} {\bibfnamefont {C.}~\bibnamefont {Thomas}}, \bibinfo {author} {\bibfnamefont {S.}~\bibnamefont {Metti}}, \bibinfo {author} {\bibfnamefont {C.~R.}\ \bibnamefont {Guinn}}, \bibinfo {author} {\bibfnamefont {R.}~\bibnamefont {Kallaher}}, \bibinfo {author} {\bibfnamefont {M.}~\bibnamefont {Lodari}}, \bibinfo {author} {\bibfnamefont {G.}~\bibnamefont {Scappucci}}, \bibinfo {author} {\bibfnamefont {T.}~\bibnamefont {Wang}}, \emph {et~al.},\ }\bibfield  {title} {\bibinfo {title} {Ballistic superconductivity and tunable $\pi$--junctions in insb quantum wells},\ }\href {https://www.nature.com/articles/s41467-019-11742-4} {\bibfield  {journal} {\bibinfo  {journal} {Nature communications}\ }\textbf {\bibinfo {volume} {10}},\ \bibinfo {pages} {1} (\bibinfo {year} {2019})}\BibitemShut
  {NoStop}%
\bibitem [{\citenamefont {Feofanov}\ \emph {et~al.}(2010)\citenamefont {Feofanov}, \citenamefont {Oboznov}, \citenamefont {Bol'ginov}, \citenamefont {Lisenfeld}, \citenamefont {Poletto}, \citenamefont {Ryazanov}, \citenamefont {Rossolenko}, \citenamefont {Khabipov}, \citenamefont {Balashov}, \citenamefont {Zorin}, \citenamefont {Dmitriev}, \citenamefont {Koshelets},\ and\ \citenamefont {Ustinov}}]{Feofanov2010}%
  \BibitemOpen
  \bibfield  {author} {\bibinfo {author} {\bibfnamefont {A.~K.}\ \bibnamefont {Feofanov}}, \bibinfo {author} {\bibfnamefont {V.~A.}\ \bibnamefont {Oboznov}}, \bibinfo {author} {\bibfnamefont {V.~V.}\ \bibnamefont {Bol'ginov}}, \bibinfo {author} {\bibfnamefont {J.}~\bibnamefont {Lisenfeld}}, \bibinfo {author} {\bibfnamefont {S.}~\bibnamefont {Poletto}}, \bibinfo {author} {\bibfnamefont {V.~V.}\ \bibnamefont {Ryazanov}}, \bibinfo {author} {\bibfnamefont {A.~N.}\ \bibnamefont {Rossolenko}}, \bibinfo {author} {\bibfnamefont {M.}~\bibnamefont {Khabipov}}, \bibinfo {author} {\bibfnamefont {D.}~\bibnamefont {Balashov}}, \bibinfo {author} {\bibfnamefont {A.~B.}\ \bibnamefont {Zorin}}, \bibinfo {author} {\bibfnamefont {P.~N.}\ \bibnamefont {Dmitriev}}, \bibinfo {author} {\bibfnamefont {V.~P.}\ \bibnamefont {Koshelets}},\ and\ \bibinfo {author} {\bibfnamefont {A.~V.}\ \bibnamefont {Ustinov}},\ }\bibfield  {title} {\bibinfo {title} {Implementation of superconductor/ferromagnet/ superconductor $\pi$-shifters in
  superconducting digital and quantum circuits},\ }\href {https://doi.org/10.1038/nphys1700} {\bibfield  {journal} {\bibinfo  {journal} {Nature Physics}\ }\textbf {\bibinfo {volume} {6}},\ \bibinfo {pages} {593} (\bibinfo {year} {2010})}\BibitemShut {NoStop}%
\bibitem [{\citenamefont {Yamashita}\ \emph {et~al.}(2005)\citenamefont {Yamashita}, \citenamefont {Tanikawa}, \citenamefont {Takahashi},\ and\ \citenamefont {Maekawa}}]{Yamashita2005}%
  \BibitemOpen
  \bibfield  {author} {\bibinfo {author} {\bibfnamefont {T.}~\bibnamefont {Yamashita}}, \bibinfo {author} {\bibfnamefont {K.}~\bibnamefont {Tanikawa}}, \bibinfo {author} {\bibfnamefont {S.}~\bibnamefont {Takahashi}},\ and\ \bibinfo {author} {\bibfnamefont {S.}~\bibnamefont {Maekawa}},\ }\bibfield  {title} {\bibinfo {title} {Superconducting $\ensuremath{\pi}$ qubit with a ferromagnetic josephson junction},\ }\href {https://doi.org/10.1103/PhysRevLett.95.097001} {\bibfield  {journal} {\bibinfo  {journal} {Phys. Rev. Lett.}\ }\textbf {\bibinfo {volume} {95}},\ \bibinfo {pages} {097001} (\bibinfo {year} {2005})}\BibitemShut {NoStop}%
\bibitem [{\citenamefont {Shcherbakova}\ \emph {et~al.}(2015)\citenamefont {Shcherbakova}, \citenamefont {Fedorov}, \citenamefont {Shulga}, \citenamefont {Ryazanov}, \citenamefont {Bolginov}, \citenamefont {Oboznov}, \citenamefont {Egorov}, \citenamefont {Shkolnikov}, \citenamefont {Wolf}, \citenamefont {Beckmann},\ and\ \citenamefont {Ustinov}}]{Shcherbakova2015}%
  \BibitemOpen
  \bibfield  {author} {\bibinfo {author} {\bibfnamefont {A.~V.}\ \bibnamefont {Shcherbakova}}, \bibinfo {author} {\bibfnamefont {K.~G.}\ \bibnamefont {Fedorov}}, \bibinfo {author} {\bibfnamefont {K.~V.}\ \bibnamefont {Shulga}}, \bibinfo {author} {\bibfnamefont {V.~V.}\ \bibnamefont {Ryazanov}}, \bibinfo {author} {\bibfnamefont {V.~V.}\ \bibnamefont {Bolginov}}, \bibinfo {author} {\bibfnamefont {V.~A.}\ \bibnamefont {Oboznov}}, \bibinfo {author} {\bibfnamefont {S.~V.}\ \bibnamefont {Egorov}}, \bibinfo {author} {\bibfnamefont {V.~O.}\ \bibnamefont {Shkolnikov}}, \bibinfo {author} {\bibfnamefont {M.~J.}\ \bibnamefont {Wolf}}, \bibinfo {author} {\bibfnamefont {D.}~\bibnamefont {Beckmann}},\ and\ \bibinfo {author} {\bibfnamefont {A.~V.}\ \bibnamefont {Ustinov}},\ }\bibfield  {title} {\bibinfo {title} {Fabrication and measurements of hybrid nb/al josephson junctions and flux qubits with $\pi$-shifters},\ }\href {https://doi.org/10.1088/0953-2048/28/2/025009} {\bibfield  {journal} {\bibinfo  {journal}
  {Superconductor Science and Technology}\ }\textbf {\bibinfo {volume} {28}},\ \bibinfo {pages} {025009} (\bibinfo {year} {2015})}\BibitemShut {NoStop}%
\bibitem [{\citenamefont {Bobkova}\ \emph {et~al.}(2022)\citenamefont {Bobkova}, \citenamefont {Bobkov},\ and\ \citenamefont {Silaev}}]{Bobkova_review}%
  \BibitemOpen
  \bibfield  {author} {\bibinfo {author} {\bibfnamefont {I.~V.}\ \bibnamefont {Bobkova}}, \bibinfo {author} {\bibfnamefont {A.~M.}\ \bibnamefont {Bobkov}},\ and\ \bibinfo {author} {\bibfnamefont {M.~A.}\ \bibnamefont {Silaev}},\ }\bibfield  {title} {\bibinfo {title} {Magnetoelectric effects in josephson junctions},\ }\href {https://doi.org/10.1088/1361-648X/ac7994} {\bibfield  {journal} {\bibinfo  {journal} {Journal of Physics: Condensed Matter}\ }\textbf {\bibinfo {volume} {34}},\ \bibinfo {pages} {353001} (\bibinfo {year} {2022})}\BibitemShut {NoStop}%
\bibitem [{\citenamefont {Shukrinov}(2022)}]{Shukrinov_review}%
  \BibitemOpen
  \bibfield  {author} {\bibinfo {author} {\bibfnamefont {Y.~M.}\ \bibnamefont {Shukrinov}},\ }\bibfield  {title} {\bibinfo {title} {Anomalous josephson effect},\ }\href {https://doi.org/10.3367/UFNe.2020.11.038894} {\bibfield  {journal} {\bibinfo  {journal} {Physics-Uspekhi}\ }\textbf {\bibinfo {volume} {65}},\ \bibinfo {pages} {317} (\bibinfo {year} {2022})}\BibitemShut {NoStop}%
\bibitem [{\citenamefont {Mayer}\ \emph {et~al.}(2020)\citenamefont {Mayer}, \citenamefont {Dartiailh}, \citenamefont {Yuan}, \citenamefont {Wickramasinghe}, \citenamefont {Rossi},\ and\ \citenamefont {Shabani}}]{Mayer2020}%
  \BibitemOpen
  \bibfield  {author} {\bibinfo {author} {\bibfnamefont {W.}~\bibnamefont {Mayer}}, \bibinfo {author} {\bibfnamefont {M.~C.}\ \bibnamefont {Dartiailh}}, \bibinfo {author} {\bibfnamefont {J.}~\bibnamefont {Yuan}}, \bibinfo {author} {\bibfnamefont {K.~S.}\ \bibnamefont {Wickramasinghe}}, \bibinfo {author} {\bibfnamefont {E.}~\bibnamefont {Rossi}},\ and\ \bibinfo {author} {\bibfnamefont {J.}~\bibnamefont {Shabani}},\ }\bibfield  {title} {\bibinfo {title} {Gate controlled anomalous phase shift in al/inas josephson junctions},\ }\href {https://doi.org/10.1038/s41467-019-14094-1} {\bibfield  {journal} {\bibinfo  {journal} {Nature Communications}\ }\textbf {\bibinfo {volume} {11}},\ \bibinfo {pages} {212} (\bibinfo {year} {2020})}\BibitemShut {NoStop}%
\bibitem [{\citenamefont {Szombati}\ \emph {et~al.}(2016)\citenamefont {Szombati}, \citenamefont {Nadj-Perge}, \citenamefont {Car}, \citenamefont {Plissard}, \citenamefont {Bakkers},\ and\ \citenamefont {Kouwenhoven}}]{Szombati2016}%
  \BibitemOpen
  \bibfield  {author} {\bibinfo {author} {\bibfnamefont {D.~B.}\ \bibnamefont {Szombati}}, \bibinfo {author} {\bibfnamefont {S.}~\bibnamefont {Nadj-Perge}}, \bibinfo {author} {\bibfnamefont {D.}~\bibnamefont {Car}}, \bibinfo {author} {\bibfnamefont {S.~R.}\ \bibnamefont {Plissard}}, \bibinfo {author} {\bibfnamefont {E.~P. A.~M.}\ \bibnamefont {Bakkers}},\ and\ \bibinfo {author} {\bibfnamefont {L.~P.}\ \bibnamefont {Kouwenhoven}},\ }\bibfield  {title} {\bibinfo {title} {Josephson $\varphi_0$-junction in nanowire quantum dots},\ }\href {https://doi.org/10.1038/nphys3742} {\bibfield  {journal} {\bibinfo  {journal} {Nature Physics}\ }\textbf {\bibinfo {volume} {12}},\ \bibinfo {pages} {568} (\bibinfo {year} {2016})}\BibitemShut {NoStop}%
\bibitem [{\citenamefont {Assouline}\ \emph {et~al.}(2019)\citenamefont {Assouline}, \citenamefont {Feuillet-Palma}, \citenamefont {Bergeal}, \citenamefont {Zhang}, \citenamefont {Mottaghizadeh}, \citenamefont {Zimmers}, \citenamefont {Lhuillier}, \citenamefont {Eddrie}, \citenamefont {Atkinson}, \citenamefont {Aprili},\ and\ \citenamefont {Aubin}}]{Assouline2019}%
  \BibitemOpen
  \bibfield  {author} {\bibinfo {author} {\bibfnamefont {A.}~\bibnamefont {Assouline}}, \bibinfo {author} {\bibfnamefont {C.}~\bibnamefont {Feuillet-Palma}}, \bibinfo {author} {\bibfnamefont {N.}~\bibnamefont {Bergeal}}, \bibinfo {author} {\bibfnamefont {T.}~\bibnamefont {Zhang}}, \bibinfo {author} {\bibfnamefont {A.}~\bibnamefont {Mottaghizadeh}}, \bibinfo {author} {\bibfnamefont {A.}~\bibnamefont {Zimmers}}, \bibinfo {author} {\bibfnamefont {E.}~\bibnamefont {Lhuillier}}, \bibinfo {author} {\bibfnamefont {M.}~\bibnamefont {Eddrie}}, \bibinfo {author} {\bibfnamefont {P.}~\bibnamefont {Atkinson}}, \bibinfo {author} {\bibfnamefont {M.}~\bibnamefont {Aprili}},\ and\ \bibinfo {author} {\bibfnamefont {H.}~\bibnamefont {Aubin}},\ }\bibfield  {title} {\bibinfo {title} {Spin-orbit induced phase-shift in bi2se3 josephson junctions},\ }\href {https://doi.org/10.1038/s41467-018-08022-y} {\bibfield  {journal} {\bibinfo  {journal} {Nature Communications}\ }\textbf {\bibinfo {volume} {10}},\ \bibinfo {pages} {126}
  (\bibinfo {year} {2019})}\BibitemShut {NoStop}%
\bibitem [{\citenamefont {Murani}\ \emph {et~al.}(2017)\citenamefont {Murani}, \citenamefont {Kasumov}, \citenamefont {Sengupta}, \citenamefont {Kasumov}, \citenamefont {Volkov}, \citenamefont {Khodos}, \citenamefont {Brisset}, \citenamefont {Delagrange}, \citenamefont {Chepelianskii}, \citenamefont {Deblock}, \citenamefont {Bouchiat},\ and\ \citenamefont {Gu{\'e}ron}}]{Murani2017}%
  \BibitemOpen
  \bibfield  {author} {\bibinfo {author} {\bibfnamefont {A.}~\bibnamefont {Murani}}, \bibinfo {author} {\bibfnamefont {A.}~\bibnamefont {Kasumov}}, \bibinfo {author} {\bibfnamefont {S.}~\bibnamefont {Sengupta}}, \bibinfo {author} {\bibfnamefont {Y.~A.}\ \bibnamefont {Kasumov}}, \bibinfo {author} {\bibfnamefont {V.~T.}\ \bibnamefont {Volkov}}, \bibinfo {author} {\bibfnamefont {I.~I.}\ \bibnamefont {Khodos}}, \bibinfo {author} {\bibfnamefont {F.}~\bibnamefont {Brisset}}, \bibinfo {author} {\bibfnamefont {R.}~\bibnamefont {Delagrange}}, \bibinfo {author} {\bibfnamefont {A.}~\bibnamefont {Chepelianskii}}, \bibinfo {author} {\bibfnamefont {R.}~\bibnamefont {Deblock}}, \bibinfo {author} {\bibfnamefont {H.}~\bibnamefont {Bouchiat}},\ and\ \bibinfo {author} {\bibfnamefont {S.}~\bibnamefont {Gu{\'e}ron}},\ }\bibfield  {title} {\bibinfo {title} {Ballistic edge states in bismuth nanowires revealed by squid interferometry},\ }\href {https://doi.org/10.1038/ncomms15941} {\bibfield  {journal} {\bibinfo  {journal} {Nature
  Communications}\ }\textbf {\bibinfo {volume} {8}},\ \bibinfo {pages} {15941} (\bibinfo {year} {2017})}\BibitemShut {NoStop}%
\bibitem [{\citenamefont {Linder}\ and\ \citenamefont {Robinson}(2015)}]{Linder2015}%
  \BibitemOpen
  \bibfield  {author} {\bibinfo {author} {\bibfnamefont {J.}~\bibnamefont {Linder}}\ and\ \bibinfo {author} {\bibfnamefont {J.~W.~A.}\ \bibnamefont {Robinson}},\ }\bibfield  {title} {\bibinfo {title} {Superconducting spintronics},\ }\href {https://doi.org/10.1038/nphys3242} {\bibfield  {journal} {\bibinfo  {journal} {Nature Physics}\ }\textbf {\bibinfo {volume} {11}},\ \bibinfo {pages} {307} (\bibinfo {year} {2015})}\BibitemShut {NoStop}%
\bibitem [{\citenamefont {Eschrig}(2015)}]{Eschrig2015}%
  \BibitemOpen
  \bibfield  {author} {\bibinfo {author} {\bibfnamefont {M.}~\bibnamefont {Eschrig}},\ }\bibfield  {title} {\bibinfo {title} {Spin-polarized supercurrents for spintronics: a review of current progress},\ }\href {https://doi.org/10.1088/0034-4885/78/10/104501} {\bibfield  {journal} {\bibinfo  {journal} {Reports on Progress in Physics}\ }\textbf {\bibinfo {volume} {78}},\ \bibinfo {pages} {104501} (\bibinfo {year} {2015})}\BibitemShut {NoStop}%
\bibitem [{\citenamefont {Konschelle}\ and\ \citenamefont {Buzdin}(2009)}]{Konschelle2009}%
  \BibitemOpen
  \bibfield  {author} {\bibinfo {author} {\bibfnamefont {F.}~\bibnamefont {Konschelle}}\ and\ \bibinfo {author} {\bibfnamefont {A.}~\bibnamefont {Buzdin}},\ }\bibfield  {title} {\bibinfo {title} {Magnetic moment manipulation by a josephson current},\ }\href {https://doi.org/10.1103/PhysRevLett.102.017001} {\bibfield  {journal} {\bibinfo  {journal} {Phys. Rev. Lett.}\ }\textbf {\bibinfo {volume} {102}},\ \bibinfo {pages} {017001} (\bibinfo {year} {2009})}\BibitemShut {NoStop}%
\bibitem [{\citenamefont {Shukrinov}\ \emph {et~al.}(2017)\citenamefont {Shukrinov}, \citenamefont {Rahmonov}, \citenamefont {Sengupta},\ and\ \citenamefont {Buzdin}}]{Shukrinov2017}%
  \BibitemOpen
  \bibfield  {author} {\bibinfo {author} {\bibfnamefont {Y.~M.}\ \bibnamefont {Shukrinov}}, \bibinfo {author} {\bibfnamefont {I.~R.}\ \bibnamefont {Rahmonov}}, \bibinfo {author} {\bibfnamefont {K.}~\bibnamefont {Sengupta}},\ and\ \bibinfo {author} {\bibfnamefont {A.}~\bibnamefont {Buzdin}},\ }\bibfield  {title} {\bibinfo {title} {Magnetization reversal by superconducting current in $\phi$ josephson junctions},\ }\href {https://doi.org/10.1063/1.4983090} {\bibfield  {journal} {\bibinfo  {journal} {Applied Physics Letters}\ }\textbf {\bibinfo {volume} {110}},\ \bibinfo {pages} {182407} (\bibinfo {year} {2017})}\BibitemShut {NoStop}%
\bibitem [{\citenamefont {Nashaat}\ \emph {et~al.}(2019)\citenamefont {Nashaat}, \citenamefont {Bobkova}, \citenamefont {Bobkov}, \citenamefont {Shukrinov}, \citenamefont {Rahmonov},\ and\ \citenamefont {Sengupta}}]{Nashaat2019}%
  \BibitemOpen
  \bibfield  {author} {\bibinfo {author} {\bibfnamefont {M.}~\bibnamefont {Nashaat}}, \bibinfo {author} {\bibfnamefont {I.~V.}\ \bibnamefont {Bobkova}}, \bibinfo {author} {\bibfnamefont {A.~M.}\ \bibnamefont {Bobkov}}, \bibinfo {author} {\bibfnamefont {Y.~M.}\ \bibnamefont {Shukrinov}}, \bibinfo {author} {\bibfnamefont {I.~R.}\ \bibnamefont {Rahmonov}},\ and\ \bibinfo {author} {\bibfnamefont {K.}~\bibnamefont {Sengupta}},\ }\bibfield  {title} {\bibinfo {title} {Electrical control of magnetization in superconductor/ferromagnet/superconductor junctions on a three-dimensional topological insulator},\ }\href {https://doi.org/10.1103/PhysRevB.100.054506} {\bibfield  {journal} {\bibinfo  {journal} {Phys. Rev. B}\ }\textbf {\bibinfo {volume} {100}},\ \bibinfo {pages} {054506} (\bibinfo {year} {2019})}\BibitemShut {NoStop}%
\bibitem [{\citenamefont {Rabinovich}\ \emph {et~al.}(2019)\citenamefont {Rabinovich}, \citenamefont {Bobkova}, \citenamefont {Bobkov},\ and\ \citenamefont {Silaev}}]{Rabinovich2019}%
  \BibitemOpen
  \bibfield  {author} {\bibinfo {author} {\bibfnamefont {D.~S.}\ \bibnamefont {Rabinovich}}, \bibinfo {author} {\bibfnamefont {I.~V.}\ \bibnamefont {Bobkova}}, \bibinfo {author} {\bibfnamefont {A.~M.}\ \bibnamefont {Bobkov}},\ and\ \bibinfo {author} {\bibfnamefont {M.~A.}\ \bibnamefont {Silaev}},\ }\bibfield  {title} {\bibinfo {title} {Resistive state of superconductor-ferromagnet-superconductor josephson junctions in the presence of moving domain walls},\ }\href {https://doi.org/10.1103/PhysRevLett.123.207001} {\bibfield  {journal} {\bibinfo  {journal} {Phys. Rev. Lett.}\ }\textbf {\bibinfo {volume} {123}},\ \bibinfo {pages} {207001} (\bibinfo {year} {2019})}\BibitemShut {NoStop}%
\bibitem [{\citenamefont {Guarcello}\ and\ \citenamefont {Bergeret}(2020)}]{Guarcello2020}%
  \BibitemOpen
  \bibfield  {author} {\bibinfo {author} {\bibfnamefont {C.}~\bibnamefont {Guarcello}}\ and\ \bibinfo {author} {\bibfnamefont {F.}~\bibnamefont {Bergeret}},\ }\bibfield  {title} {\bibinfo {title} {Cryogenic memory element based on an anomalous josephson junction},\ }\href {https://doi.org/10.1103/PhysRevApplied.13.034012} {\bibfield  {journal} {\bibinfo  {journal} {Phys. Rev. Appl.}\ }\textbf {\bibinfo {volume} {13}},\ \bibinfo {pages} {034012} (\bibinfo {year} {2020})}\BibitemShut {NoStop}%
\bibitem [{\citenamefont {Bobkova}\ \emph {et~al.}(2020)\citenamefont {Bobkova}, \citenamefont {Bobkov}, \citenamefont {Rahmonov}, \citenamefont {Mazanik}, \citenamefont {Sengupta},\ and\ \citenamefont {Shukrinov}}]{Bobkova2020}%
  \BibitemOpen
  \bibfield  {author} {\bibinfo {author} {\bibfnamefont {I.~V.}\ \bibnamefont {Bobkova}}, \bibinfo {author} {\bibfnamefont {A.~M.}\ \bibnamefont {Bobkov}}, \bibinfo {author} {\bibfnamefont {I.~R.}\ \bibnamefont {Rahmonov}}, \bibinfo {author} {\bibfnamefont {A.~A.}\ \bibnamefont {Mazanik}}, \bibinfo {author} {\bibfnamefont {K.}~\bibnamefont {Sengupta}},\ and\ \bibinfo {author} {\bibfnamefont {Y.~M.}\ \bibnamefont {Shukrinov}},\ }\bibfield  {title} {\bibinfo {title} {Magnetization reversal in superconductor/insulating ferromagnet/superconductor josephson junctions on a three-dimensional topological insulator},\ }\href {https://doi.org/10.1103/PhysRevB.102.134505} {\bibfield  {journal} {\bibinfo  {journal} {Phys. Rev. B}\ }\textbf {\bibinfo {volume} {102}},\ \bibinfo {pages} {134505} (\bibinfo {year} {2020})}\BibitemShut {NoStop}%
\bibitem [{\citenamefont {Chang}\ \emph {et~al.}(2013)\citenamefont {Chang}, \citenamefont {Zhang}, \citenamefont {Liu}, \citenamefont {Zhang}, \citenamefont {Feng}, \citenamefont {Li}, \citenamefont {Wang}, \citenamefont {Chen}, \citenamefont {Dai}, \citenamefont {Fang}, \citenamefont {Qi}, \citenamefont {Zhang}, \citenamefont {Wang}, \citenamefont {He}, \citenamefont {Ma},\ and\ \citenamefont {Xue}}]{Chang2013}%
  \BibitemOpen
  \bibfield  {author} {\bibinfo {author} {\bibfnamefont {C.-Z.}\ \bibnamefont {Chang}}, \bibinfo {author} {\bibfnamefont {J.}~\bibnamefont {Zhang}}, \bibinfo {author} {\bibfnamefont {M.}~\bibnamefont {Liu}}, \bibinfo {author} {\bibfnamefont {Z.}~\bibnamefont {Zhang}}, \bibinfo {author} {\bibfnamefont {X.}~\bibnamefont {Feng}}, \bibinfo {author} {\bibfnamefont {K.}~\bibnamefont {Li}}, \bibinfo {author} {\bibfnamefont {L.-L.}\ \bibnamefont {Wang}}, \bibinfo {author} {\bibfnamefont {X.}~\bibnamefont {Chen}}, \bibinfo {author} {\bibfnamefont {X.}~\bibnamefont {Dai}}, \bibinfo {author} {\bibfnamefont {Z.}~\bibnamefont {Fang}}, \bibinfo {author} {\bibfnamefont {X.-L.}\ \bibnamefont {Qi}}, \bibinfo {author} {\bibfnamefont {S.-C.}\ \bibnamefont {Zhang}}, \bibinfo {author} {\bibfnamefont {Y.}~\bibnamefont {Wang}}, \bibinfo {author} {\bibfnamefont {K.}~\bibnamefont {He}}, \bibinfo {author} {\bibfnamefont {X.-C.}\ \bibnamefont {Ma}},\ and\ \bibinfo {author} {\bibfnamefont {Q.-K.}\ \bibnamefont {Xue}},\ }\bibfield
  {title} {\bibinfo {title} {Thin films of magnetically doped topological insulator with carrier-independent long-range ferromagnetic order},\ }\href {https://doi.org/https://doi.org/10.1002/adma.201203493} {\bibfield  {journal} {\bibinfo  {journal} {Advanced Materials}\ }\textbf {\bibinfo {volume} {25}},\ \bibinfo {pages} {1065} (\bibinfo {year} {2013})}\BibitemShut {NoStop}%
\bibitem [{\citenamefont {Kou}\ \emph {et~al.}(2013{\natexlab{a}})\citenamefont {Kou}, \citenamefont {Lang}, \citenamefont {Fan}, \citenamefont {Jiang}, \citenamefont {Nie}, \citenamefont {Zhang}, \citenamefont {Jiang}, \citenamefont {Wang}, \citenamefont {Yao}, \citenamefont {He},\ and\ \citenamefont {Wang}}]{Kou2013}%
  \BibitemOpen
  \bibfield  {author} {\bibinfo {author} {\bibfnamefont {X.}~\bibnamefont {Kou}}, \bibinfo {author} {\bibfnamefont {M.}~\bibnamefont {Lang}}, \bibinfo {author} {\bibfnamefont {Y.}~\bibnamefont {Fan}}, \bibinfo {author} {\bibfnamefont {Y.}~\bibnamefont {Jiang}}, \bibinfo {author} {\bibfnamefont {T.}~\bibnamefont {Nie}}, \bibinfo {author} {\bibfnamefont {J.}~\bibnamefont {Zhang}}, \bibinfo {author} {\bibfnamefont {W.}~\bibnamefont {Jiang}}, \bibinfo {author} {\bibfnamefont {Y.}~\bibnamefont {Wang}}, \bibinfo {author} {\bibfnamefont {Y.}~\bibnamefont {Yao}}, \bibinfo {author} {\bibfnamefont {L.}~\bibnamefont {He}},\ and\ \bibinfo {author} {\bibfnamefont {K.~L.}\ \bibnamefont {Wang}},\ }\bibfield  {title} {\bibinfo {title} {Interplay between different magnetisms in cr-doped topological insulators},\ }\href {https://doi.org/10.1021/nn4038145} {\bibfield  {journal} {\bibinfo  {journal} {ACS Nano}\ }\textbf {\bibinfo {volume} {7}},\ \bibinfo {pages} {9205} (\bibinfo {year} {2013}{\natexlab{a}})}\BibitemShut
  {NoStop}%
\bibitem [{\citenamefont {Kou}\ \emph {et~al.}(2013{\natexlab{b}})\citenamefont {Kou}, \citenamefont {He}, \citenamefont {Lang}, \citenamefont {Fan}, \citenamefont {Wong}, \citenamefont {Jiang}, \citenamefont {Nie}, \citenamefont {Jiang}, \citenamefont {Upadhyaya}, \citenamefont {Xing}, \citenamefont {Wang}, \citenamefont {Xiu}, \citenamefont {Schwartz},\ and\ \citenamefont {Wang}}]{Kou2013_2}%
  \BibitemOpen
  \bibfield  {author} {\bibinfo {author} {\bibfnamefont {X.}~\bibnamefont {Kou}}, \bibinfo {author} {\bibfnamefont {L.}~\bibnamefont {He}}, \bibinfo {author} {\bibfnamefont {M.}~\bibnamefont {Lang}}, \bibinfo {author} {\bibfnamefont {Y.}~\bibnamefont {Fan}}, \bibinfo {author} {\bibfnamefont {K.}~\bibnamefont {Wong}}, \bibinfo {author} {\bibfnamefont {Y.}~\bibnamefont {Jiang}}, \bibinfo {author} {\bibfnamefont {T.}~\bibnamefont {Nie}}, \bibinfo {author} {\bibfnamefont {W.}~\bibnamefont {Jiang}}, \bibinfo {author} {\bibfnamefont {P.}~\bibnamefont {Upadhyaya}}, \bibinfo {author} {\bibfnamefont {Z.}~\bibnamefont {Xing}}, \bibinfo {author} {\bibfnamefont {Y.}~\bibnamefont {Wang}}, \bibinfo {author} {\bibfnamefont {F.}~\bibnamefont {Xiu}}, \bibinfo {author} {\bibfnamefont {R.~N.}\ \bibnamefont {Schwartz}},\ and\ \bibinfo {author} {\bibfnamefont {K.~L.}\ \bibnamefont {Wang}},\ }\bibfield  {title} {\bibinfo {title} {Manipulating surface-related ferromagnetism in modulation-doped topological insulators},\ }\href
  {https://doi.org/10.1021/nl4020638} {\bibfield  {journal} {\bibinfo  {journal} {Nano Letters}\ }\textbf {\bibinfo {volume} {13}},\ \bibinfo {pages} {4587} (\bibinfo {year} {2013}{\natexlab{b}})}\BibitemShut {NoStop}%
\bibitem [{\citenamefont {Chang}\ \emph {et~al.}(2015)\citenamefont {Chang}, \citenamefont {Zhao}, \citenamefont {Kim}, \citenamefont {Zhang}, \citenamefont {Assaf}, \citenamefont {Heiman}, \citenamefont {Zhang}, \citenamefont {Liu}, \citenamefont {Chan},\ and\ \citenamefont {Moodera}}]{Chang2015}%
  \BibitemOpen
  \bibfield  {author} {\bibinfo {author} {\bibfnamefont {C.-Z.}\ \bibnamefont {Chang}}, \bibinfo {author} {\bibfnamefont {W.}~\bibnamefont {Zhao}}, \bibinfo {author} {\bibfnamefont {D.~Y.}\ \bibnamefont {Kim}}, \bibinfo {author} {\bibfnamefont {H.}~\bibnamefont {Zhang}}, \bibinfo {author} {\bibfnamefont {B.~A.}\ \bibnamefont {Assaf}}, \bibinfo {author} {\bibfnamefont {D.}~\bibnamefont {Heiman}}, \bibinfo {author} {\bibfnamefont {S.-C.}\ \bibnamefont {Zhang}}, \bibinfo {author} {\bibfnamefont {C.}~\bibnamefont {Liu}}, \bibinfo {author} {\bibfnamefont {M.~H.~W.}\ \bibnamefont {Chan}},\ and\ \bibinfo {author} {\bibfnamefont {J.~S.}\ \bibnamefont {Moodera}},\ }\bibfield  {title} {\bibinfo {title} {High-precision realization of robust quantum anomalous hall state in a hard ferromagnetic topological insulator},\ }\href {https://doi.org/10.1038/nmat4204} {\bibfield  {journal} {\bibinfo  {journal} {Nature Materials}\ }\textbf {\bibinfo {volume} {14}},\ \bibinfo {pages} {473} (\bibinfo {year} {2015})}\BibitemShut
  {NoStop}%
\bibitem [{\citenamefont {Jiang}\ \emph {et~al.}(2014)\citenamefont {Jiang}, \citenamefont {Katmis}, \citenamefont {Tang}, \citenamefont {Wei}, \citenamefont {Moodera},\ and\ \citenamefont {Shi}}]{Jiang2014}%
  \BibitemOpen
  \bibfield  {author} {\bibinfo {author} {\bibfnamefont {Z.}~\bibnamefont {Jiang}}, \bibinfo {author} {\bibfnamefont {F.}~\bibnamefont {Katmis}}, \bibinfo {author} {\bibfnamefont {C.}~\bibnamefont {Tang}}, \bibinfo {author} {\bibfnamefont {P.}~\bibnamefont {Wei}}, \bibinfo {author} {\bibfnamefont {J.~S.}\ \bibnamefont {Moodera}},\ and\ \bibinfo {author} {\bibfnamefont {J.}~\bibnamefont {Shi}},\ }\bibfield  {title} {\bibinfo {title} {A comparative transport study of bi2se3 and bi2se3/yttrium iron garnet},\ }\href {https://doi.org/10.1063/1.4881975} {\bibfield  {journal} {\bibinfo  {journal} {Applied Physics Letters}\ }\textbf {\bibinfo {volume} {104}},\ \bibinfo {pages} {222409} (\bibinfo {year} {2014})}\BibitemShut {NoStop}%
\bibitem [{\citenamefont {Wei}\ \emph {et~al.}(2013)\citenamefont {Wei}, \citenamefont {Katmis}, \citenamefont {Assaf}, \citenamefont {Steinberg}, \citenamefont {Jarillo-Herrero}, \citenamefont {Heiman},\ and\ \citenamefont {Moodera}}]{Wei2013}%
  \BibitemOpen
  \bibfield  {author} {\bibinfo {author} {\bibfnamefont {P.}~\bibnamefont {Wei}}, \bibinfo {author} {\bibfnamefont {F.}~\bibnamefont {Katmis}}, \bibinfo {author} {\bibfnamefont {B.~A.}\ \bibnamefont {Assaf}}, \bibinfo {author} {\bibfnamefont {H.}~\bibnamefont {Steinberg}}, \bibinfo {author} {\bibfnamefont {P.}~\bibnamefont {Jarillo-Herrero}}, \bibinfo {author} {\bibfnamefont {D.}~\bibnamefont {Heiman}},\ and\ \bibinfo {author} {\bibfnamefont {J.~S.}\ \bibnamefont {Moodera}},\ }\bibfield  {title} {\bibinfo {title} {Exchange-coupling-induced symmetry breaking in topological insulators},\ }\href {https://doi.org/10.1103/PhysRevLett.110.186807} {\bibfield  {journal} {\bibinfo  {journal} {Phys. Rev. Lett.}\ }\textbf {\bibinfo {volume} {110}},\ \bibinfo {pages} {186807} (\bibinfo {year} {2013})}\BibitemShut {NoStop}%
\bibitem [{\citenamefont {Jiang}\ \emph {et~al.}(2015)\citenamefont {Jiang}, \citenamefont {Chang}, \citenamefont {Tang}, \citenamefont {Wei}, \citenamefont {Moodera},\ and\ \citenamefont {Shi}}]{Jiang2015}%
  \BibitemOpen
  \bibfield  {author} {\bibinfo {author} {\bibfnamefont {Z.}~\bibnamefont {Jiang}}, \bibinfo {author} {\bibfnamefont {C.-Z.}\ \bibnamefont {Chang}}, \bibinfo {author} {\bibfnamefont {C.}~\bibnamefont {Tang}}, \bibinfo {author} {\bibfnamefont {P.}~\bibnamefont {Wei}}, \bibinfo {author} {\bibfnamefont {J.~S.}\ \bibnamefont {Moodera}},\ and\ \bibinfo {author} {\bibfnamefont {J.}~\bibnamefont {Shi}},\ }\bibfield  {title} {\bibinfo {title} {Independent tuning of electronic properties and induced ferromagnetism in topological insulators with heterostructure approach},\ }\href {https://doi.org/10.1021/acs.nanolett.5b01905} {\bibfield  {journal} {\bibinfo  {journal} {Nano Letters}\ }\textbf {\bibinfo {volume} {15}},\ \bibinfo {pages} {5835} (\bibinfo {year} {2015})}\BibitemShut {NoStop}%
\bibitem [{\citenamefont {Jiang}\ \emph {et~al.}(2016)\citenamefont {Jiang}, \citenamefont {Chang}, \citenamefont {Tang}, \citenamefont {Zheng}, \citenamefont {Moodera},\ and\ \citenamefont {Shi}}]{Jiang2016}%
  \BibitemOpen
  \bibfield  {author} {\bibinfo {author} {\bibfnamefont {Z.}~\bibnamefont {Jiang}}, \bibinfo {author} {\bibfnamefont {C.-Z.}\ \bibnamefont {Chang}}, \bibinfo {author} {\bibfnamefont {C.}~\bibnamefont {Tang}}, \bibinfo {author} {\bibfnamefont {J.-G.}\ \bibnamefont {Zheng}}, \bibinfo {author} {\bibfnamefont {J.~S.}\ \bibnamefont {Moodera}},\ and\ \bibinfo {author} {\bibfnamefont {J.}~\bibnamefont {Shi}},\ }\bibfield  {title} {\bibinfo {title} {Structural and proximity-induced ferromagnetic properties of topological insulator-magnetic insulator heterostructures},\ }\href {https://doi.org/10.1063/1.4943061} {\bibfield  {journal} {\bibinfo  {journal} {AIP Advances}\ }\textbf {\bibinfo {volume} {6}},\ \bibinfo {pages} {055809} (\bibinfo {year} {2016})}\BibitemShut {NoStop}%
\bibitem [{\citenamefont {Gibertini}\ \emph {et~al.}(2019)\citenamefont {Gibertini}, \citenamefont {Koperski}, \citenamefont {Morpurgo},\ and\ \citenamefont {Novoselov}}]{Gibertini2019}%
  \BibitemOpen
  \bibfield  {author} {\bibinfo {author} {\bibfnamefont {M.}~\bibnamefont {Gibertini}}, \bibinfo {author} {\bibfnamefont {M.}~\bibnamefont {Koperski}}, \bibinfo {author} {\bibfnamefont {A.~F.}\ \bibnamefont {Morpurgo}},\ and\ \bibinfo {author} {\bibfnamefont {K.~S.}\ \bibnamefont {Novoselov}},\ }\bibfield  {title} {\bibinfo {title} {Magnetic 2d materials and heterostructures},\ }\href {https://doi.org/10.1038/s41565-019-0438-6} {\bibfield  {journal} {\bibinfo  {journal} {Nature Nanotechnology}\ }\textbf {\bibinfo {volume} {14}},\ \bibinfo {pages} {408} (\bibinfo {year} {2019})}\BibitemShut {NoStop}%
\bibitem [{\citenamefont {Ai}\ \emph {et~al.}(2021)\citenamefont {Ai}, \citenamefont {Zhang}, \citenamefont {Yang}, \citenamefont {Xie}, \citenamefont {Yang}, \citenamefont {Jia}, \citenamefont {Zhang}, \citenamefont {Liu}, \citenamefont {Li}, \citenamefont {Leng}, \citenamefont {Cao}, \citenamefont {Sun}, \citenamefont {Zhang}, \citenamefont {Kou}, \citenamefont {Han}, \citenamefont {Xiu},\ and\ \citenamefont {Dong}}]{Ai2021}%
  \BibitemOpen
  \bibfield  {author} {\bibinfo {author} {\bibfnamefont {L.}~\bibnamefont {Ai}}, \bibinfo {author} {\bibfnamefont {E.}~\bibnamefont {Zhang}}, \bibinfo {author} {\bibfnamefont {J.}~\bibnamefont {Yang}}, \bibinfo {author} {\bibfnamefont {X.}~\bibnamefont {Xie}}, \bibinfo {author} {\bibfnamefont {Y.}~\bibnamefont {Yang}}, \bibinfo {author} {\bibfnamefont {Z.}~\bibnamefont {Jia}}, \bibinfo {author} {\bibfnamefont {Y.}~\bibnamefont {Zhang}}, \bibinfo {author} {\bibfnamefont {S.}~\bibnamefont {Liu}}, \bibinfo {author} {\bibfnamefont {Z.}~\bibnamefont {Li}}, \bibinfo {author} {\bibfnamefont {P.}~\bibnamefont {Leng}}, \bibinfo {author} {\bibfnamefont {X.}~\bibnamefont {Cao}}, \bibinfo {author} {\bibfnamefont {X.}~\bibnamefont {Sun}}, \bibinfo {author} {\bibfnamefont {T.}~\bibnamefont {Zhang}}, \bibinfo {author} {\bibfnamefont {X.}~\bibnamefont {Kou}}, \bibinfo {author} {\bibfnamefont {Z.}~\bibnamefont {Han}}, \bibinfo {author} {\bibfnamefont {F.}~\bibnamefont {Xiu}},\ and\ \bibinfo {author} {\bibfnamefont
  {S.}~\bibnamefont {Dong}},\ }\bibfield  {title} {\bibinfo {title} {Van der waals ferromagnetic josephson junctions},\ }\href {https://doi.org/10.1038/s41467-021-26946-w} {\bibfield  {journal} {\bibinfo  {journal} {Nature Communications}\ }\textbf {\bibinfo {volume} {12}},\ \bibinfo {pages} {6580} (\bibinfo {year} {2021})}\BibitemShut {NoStop}%
\bibitem [{\citenamefont {Kang}\ \emph {et~al.}(2022)\citenamefont {Kang}, \citenamefont {Berger}, \citenamefont {Watanabe}, \citenamefont {Taniguchi}, \citenamefont {Forr{\'o}}, \citenamefont {Shan},\ and\ \citenamefont {Mak}}]{Kang2022}%
  \BibitemOpen
  \bibfield  {author} {\bibinfo {author} {\bibfnamefont {K.}~\bibnamefont {Kang}}, \bibinfo {author} {\bibfnamefont {H.}~\bibnamefont {Berger}}, \bibinfo {author} {\bibfnamefont {K.}~\bibnamefont {Watanabe}}, \bibinfo {author} {\bibfnamefont {T.}~\bibnamefont {Taniguchi}}, \bibinfo {author} {\bibfnamefont {L.}~\bibnamefont {Forr{\'o}}}, \bibinfo {author} {\bibfnamefont {J.}~\bibnamefont {Shan}},\ and\ \bibinfo {author} {\bibfnamefont {K.~F.}\ \bibnamefont {Mak}},\ }\bibfield  {title} {\bibinfo {title} {van der waals $\pi$ josephson junctions},\ }\href {https://doi.org/10.1021/acs.nanolett.2c01640} {\bibfield  {journal} {\bibinfo  {journal} {Nano Letters}\ }\textbf {\bibinfo {volume} {22}},\ \bibinfo {pages} {5510} (\bibinfo {year} {2022})}\BibitemShut {NoStop}%
\bibitem [{\citenamefont {Bobkov}\ \emph {et~al.}(2022)\citenamefont {Bobkov}, \citenamefont {Bobkova},\ and\ \citenamefont {Bobkov}}]{Bobkov2022}%
  \BibitemOpen
  \bibfield  {author} {\bibinfo {author} {\bibfnamefont {G.~A.}\ \bibnamefont {Bobkov}}, \bibinfo {author} {\bibfnamefont {I.~V.}\ \bibnamefont {Bobkova}},\ and\ \bibinfo {author} {\bibfnamefont {A.~M.}\ \bibnamefont {Bobkov}},\ }\bibfield  {title} {\bibinfo {title} {Long-range interaction of magnetic moments in a coupled system of superconductor-ferromagnet-superconductor josephson junctions with anomalous ground-state phase shift},\ }\href {https://doi.org/10.1103/PhysRevB.105.024513} {\bibfield  {journal} {\bibinfo  {journal} {Phys. Rev. B}\ }\textbf {\bibinfo {volume} {105}},\ \bibinfo {pages} {024513} (\bibinfo {year} {2022})}\BibitemShut {NoStop}%
\bibitem [{\citenamefont {Bobkov}\ \emph {et~al.}(2024{\natexlab{a}})\citenamefont {Bobkov}, \citenamefont {Bobkova},\ and\ \citenamefont {Bobkov}}]{Bobkov2024modes}%
  \BibitemOpen
  \bibfield  {author} {\bibinfo {author} {\bibfnamefont {G.~A.}\ \bibnamefont {Bobkov}}, \bibinfo {author} {\bibfnamefont {I.~V.}\ \bibnamefont {Bobkova}},\ and\ \bibinfo {author} {\bibfnamefont {A.~M.}\ \bibnamefont {Bobkov}},\ }\bibfield  {title} {\bibinfo {title} {Magnetic eigenmodes in chains of coupled $\varphi_0$-josephson junctions with ferromagnetic weak links},\ }\href {https://doi.org/10.1134/S0021364023604013} {\bibfield  {journal} {\bibinfo  {journal} {JETP Lett.}\ } (\bibinfo {year} {2024}{\natexlab{a}})}\BibitemShut {NoStop}%
\bibitem [{\citenamefont {Bobkov}\ \emph {et~al.}(2024{\natexlab{b}})\citenamefont {Bobkov}, \citenamefont {Bobkova},\ and\ \citenamefont {Bobkov}}]{Bobkov2024many}%
  \BibitemOpen
  \bibfield  {author} {\bibinfo {author} {\bibfnamefont {G.~A.}\ \bibnamefont {Bobkov}}, \bibinfo {author} {\bibfnamefont {I.~V.}\ \bibnamefont {Bobkova}},\ and\ \bibinfo {author} {\bibfnamefont {A.~M.}\ \bibnamefont {Bobkov}},\ }\bibfield  {title} {\bibinfo {title} {Controllable magnetic states in chains of coupled ${\ensuremath{\varphi}}_{0}$ josephson junctions with ferromagnetic weak links},\ }\href {https://doi.org/10.1103/PhysRevB.109.054523} {\bibfield  {journal} {\bibinfo  {journal} {Phys. Rev. B}\ }\textbf {\bibinfo {volume} {109}},\ \bibinfo {pages} {054523} (\bibinfo {year} {2024}{\natexlab{b}})}\BibitemShut {NoStop}%
\bibitem [{\citenamefont {Buzdin}(2008)}]{Buzdin2008}%
  \BibitemOpen
  \bibfield  {author} {\bibinfo {author} {\bibfnamefont {A.}~\bibnamefont {Buzdin}},\ }\bibfield  {title} {\bibinfo {title} {Direct coupling between magnetism and superconducting current in the josephson ${\ensuremath{\varphi}}_{0}$ junction},\ }\href {https://doi.org/10.1103/PhysRevLett.101.107005} {\bibfield  {journal} {\bibinfo  {journal} {Phys. Rev. Lett.}\ }\textbf {\bibinfo {volume} {101}},\ \bibinfo {pages} {107005} (\bibinfo {year} {2008})}\BibitemShut {NoStop}%
\bibitem [{\citenamefont {{Bergeret, F. S.}}\ and\ \citenamefont {{Tokatly, I. V.}}(2015)}]{Bergeret2015}%
  \BibitemOpen
  \bibfield  {author} {\bibinfo {author} {\bibnamefont {{Bergeret, F. S.}}}\ and\ \bibinfo {author} {\bibnamefont {{Tokatly, I. V.}}},\ }\bibfield  {title} {\bibinfo {title} {Theory of diffusive sephson junctions in the presence of spin-orbit coupling},\ }\href {https://doi.org/10.1209/0295-5075/110/57005} {\bibfield  {journal} {\bibinfo  {journal} {EPL}\ }\textbf {\bibinfo {volume} {110}},\ \bibinfo {pages} {57005} (\bibinfo {year} {2015})}\BibitemShut {NoStop}%
\bibitem [{\citenamefont {Zyuzin}\ \emph {et~al.}(2016)\citenamefont {Zyuzin}, \citenamefont {Alidoust},\ and\ \citenamefont {Loss}}]{Zyuzin2016}%
  \BibitemOpen
  \bibfield  {author} {\bibinfo {author} {\bibfnamefont {A.}~\bibnamefont {Zyuzin}}, \bibinfo {author} {\bibfnamefont {M.}~\bibnamefont {Alidoust}},\ and\ \bibinfo {author} {\bibfnamefont {D.}~\bibnamefont {Loss}},\ }\bibfield  {title} {\bibinfo {title} {Josephson junction through a disordered topological insulator with helical magnetization},\ }\href {https://doi.org/10.1103/PhysRevB.93.214502} {\bibfield  {journal} {\bibinfo  {journal} {Phys. Rev. B}\ }\textbf {\bibinfo {volume} {93}},\ \bibinfo {pages} {214502} (\bibinfo {year} {2016})}\BibitemShut {NoStop}%
\bibitem [{\citenamefont {Veldhorst}\ \emph {et~al.}(2012)\citenamefont {Veldhorst}, \citenamefont {Snelder}, \citenamefont {Hoek}, \citenamefont {Gang}, \citenamefont {Guduru}, \citenamefont {Wang}, \citenamefont {Zeitler}, \citenamefont {van~der Wiel}, \citenamefont {Golubov}, \citenamefont {Hilgenkamp},\ and\ \citenamefont {Brinkman}}]{Veldhorst2012}%
  \BibitemOpen
  \bibfield  {author} {\bibinfo {author} {\bibfnamefont {M.}~\bibnamefont {Veldhorst}}, \bibinfo {author} {\bibfnamefont {M.}~\bibnamefont {Snelder}}, \bibinfo {author} {\bibfnamefont {M.}~\bibnamefont {Hoek}}, \bibinfo {author} {\bibfnamefont {T.}~\bibnamefont {Gang}}, \bibinfo {author} {\bibfnamefont {V.~K.}\ \bibnamefont {Guduru}}, \bibinfo {author} {\bibfnamefont {X.~L.}\ \bibnamefont {Wang}}, \bibinfo {author} {\bibfnamefont {U.}~\bibnamefont {Zeitler}}, \bibinfo {author} {\bibfnamefont {W.~G.}\ \bibnamefont {van~der Wiel}}, \bibinfo {author} {\bibfnamefont {A.~A.}\ \bibnamefont {Golubov}}, \bibinfo {author} {\bibfnamefont {H.}~\bibnamefont {Hilgenkamp}},\ and\ \bibinfo {author} {\bibfnamefont {A.}~\bibnamefont {Brinkman}},\ }\bibfield  {title} {\bibinfo {title} {Josephson supercurrent through a topological insulator surface state},\ }\href {https://doi.org/10.1038/nmat3255} {\bibfield  {journal} {\bibinfo  {journal} {Nature Materials}\ }\textbf {\bibinfo {volume} {11}},\ \bibinfo {pages} {417} (\bibinfo
  {year} {2012})}\BibitemShut {NoStop}%
\bibitem [{\citenamefont {Xiao}\ \emph {et~al.}(2010)\citenamefont {Xiao}, \citenamefont {Bauer}, \citenamefont {Uchida}, \citenamefont {Saitoh},\ and\ \citenamefont {Maekawa}}]{Xiao2010}%
  \BibitemOpen
  \bibfield  {author} {\bibinfo {author} {\bibfnamefont {J.}~\bibnamefont {Xiao}}, \bibinfo {author} {\bibfnamefont {G.~E.~W.}\ \bibnamefont {Bauer}}, \bibinfo {author} {\bibfnamefont {K.-c.}\ \bibnamefont {Uchida}}, \bibinfo {author} {\bibfnamefont {E.}~\bibnamefont {Saitoh}},\ and\ \bibinfo {author} {\bibfnamefont {S.}~\bibnamefont {Maekawa}},\ }\bibfield  {title} {\bibinfo {title} {Theory of magnon-driven spin seebeck effect},\ }\href {https://doi.org/10.1103/PhysRevB.81.214418} {\bibfield  {journal} {\bibinfo  {journal} {Phys. Rev. B}\ }\textbf {\bibinfo {volume} {81}},\ \bibinfo {pages} {214418} (\bibinfo {year} {2010})}\BibitemShut {NoStop}%
\bibitem [{\citenamefont {Yokoyama}(2011)}]{Yokoyama2011}%
  \BibitemOpen
  \bibfield  {author} {\bibinfo {author} {\bibfnamefont {T.}~\bibnamefont {Yokoyama}},\ }\bibfield  {title} {\bibinfo {title} {Current-induced magnetization reversal on the surface of a topological insulator},\ }\href {https://doi.org/10.1103/PhysRevB.84.113407} {\bibfield  {journal} {\bibinfo  {journal} {Phys. Rev. B}\ }\textbf {\bibinfo {volume} {84}},\ \bibinfo {pages} {113407} (\bibinfo {year} {2011})}\BibitemShut {NoStop}%
\bibitem [{\citenamefont {Mihai~Miron}\ \emph {et~al.}(2010)\citenamefont {Mihai~Miron}, \citenamefont {Gaudin}, \citenamefont {Auffret}, \citenamefont {Rodmacq}, \citenamefont {Schuhl}, \citenamefont {Pizzini}, \citenamefont {Vogel},\ and\ \citenamefont {Gambardella}}]{Miron2010}%
  \BibitemOpen
  \bibfield  {author} {\bibinfo {author} {\bibfnamefont {I.}~\bibnamefont {Mihai~Miron}}, \bibinfo {author} {\bibfnamefont {G.}~\bibnamefont {Gaudin}}, \bibinfo {author} {\bibfnamefont {S.}~\bibnamefont {Auffret}}, \bibinfo {author} {\bibfnamefont {B.}~\bibnamefont {Rodmacq}}, \bibinfo {author} {\bibfnamefont {A.}~\bibnamefont {Schuhl}}, \bibinfo {author} {\bibfnamefont {S.}~\bibnamefont {Pizzini}}, \bibinfo {author} {\bibfnamefont {J.}~\bibnamefont {Vogel}},\ and\ \bibinfo {author} {\bibfnamefont {P.}~\bibnamefont {Gambardella}},\ }\bibfield  {title} {\bibinfo {title} {Current-driven spin torque induced by the rashba effect in a ferromagnetic metal layer},\ }\href {https://doi.org/10.1038/nmat2613} {\bibfield  {journal} {\bibinfo  {journal} {Nature Materials}\ }\textbf {\bibinfo {volume} {9}},\ \bibinfo {pages} {230} (\bibinfo {year} {2010})}\BibitemShut {NoStop}%
\bibitem [{\citenamefont {Bobkova}\ \emph {et~al.}(2018)\citenamefont {Bobkova}, \citenamefont {Bobkov},\ and\ \citenamefont {Silaev}}]{Bobkova2018}%
  \BibitemOpen
  \bibfield  {author} {\bibinfo {author} {\bibfnamefont {I.~V.}\ \bibnamefont {Bobkova}}, \bibinfo {author} {\bibfnamefont {A.~M.}\ \bibnamefont {Bobkov}},\ and\ \bibinfo {author} {\bibfnamefont {M.~A.}\ \bibnamefont {Silaev}},\ }\bibfield  {title} {\bibinfo {title} {Spin torques and magnetic texture dynamics driven by the supercurrent in superconductor/ferromagnet structures},\ }\href {https://doi.org/10.1103/PhysRevB.98.014521} {\bibfield  {journal} {\bibinfo  {journal} {Phys. Rev. B}\ }\textbf {\bibinfo {volume} {98}},\ \bibinfo {pages} {014521} (\bibinfo {year} {2018})}\BibitemShut {NoStop}%
\bibitem [{\citenamefont {Mazanik}\ \emph {et~al.}(2023)\citenamefont {Mazanik}, \citenamefont {Botha}, \citenamefont {Rahmonov},\ and\ \citenamefont {Shukrinov}}]{Mazanik2024}%
  \BibitemOpen
  \bibfield  {author} {\bibinfo {author} {\bibfnamefont {A.~A.}\ \bibnamefont {Mazanik}}, \bibinfo {author} {\bibfnamefont {A.~E.}\ \bibnamefont {Botha}}, \bibinfo {author} {\bibfnamefont {I.~R.}\ \bibnamefont {Rahmonov}},\ and\ \bibinfo {author} {\bibfnamefont {Y.~M.}\ \bibnamefont {Shukrinov}},\ }\href@noop {} {\bibinfo {title} {Hysteresis and chaos in anomalous josephson junctions without capacitance}} (\bibinfo {year} {2023}),\ \Eprint {https://arxiv.org/abs/2311.00597} {arXiv:2311.00597 [cond-mat.supr-con]} \BibitemShut {NoStop}%
\bibitem [{\citenamefont {Vodolazov}\ \emph {et~al.}(2003)\citenamefont {Vodolazov}, \citenamefont {Peeters}, \citenamefont {Piraux}, \citenamefont {M\'at\'efi-Tempfli},\ and\ \citenamefont {Michotte}}]{Vodolazov2003}%
  \BibitemOpen
  \bibfield  {author} {\bibinfo {author} {\bibfnamefont {D.~Y.}\ \bibnamefont {Vodolazov}}, \bibinfo {author} {\bibfnamefont {F.~M.}\ \bibnamefont {Peeters}}, \bibinfo {author} {\bibfnamefont {L.}~\bibnamefont {Piraux}}, \bibinfo {author} {\bibfnamefont {S.}~\bibnamefont {M\'at\'efi-Tempfli}},\ and\ \bibinfo {author} {\bibfnamefont {S.}~\bibnamefont {Michotte}},\ }\bibfield  {title} {\bibinfo {title} {Current-voltage characteristics of quasi-one-dimensional superconductors: An $s$-shaped curve in the constant voltage regime},\ }\href {https://doi.org/10.1103/PhysRevLett.91.157001} {\bibfield  {journal} {\bibinfo  {journal} {Phys. Rev. Lett.}\ }\textbf {\bibinfo {volume} {91}},\ \bibinfo {pages} {157001} (\bibinfo {year} {2003})}\BibitemShut {NoStop}%
\bibitem [{\citenamefont {Pop}\ \emph {et~al.}(2010)\citenamefont {Pop}, \citenamefont {Protopopov}, \citenamefont {Lecocq}, \citenamefont {Peng}, \citenamefont {Pannetier}, \citenamefont {Buisson},\ and\ \citenamefont {Guichard}}]{Pop2010}%
  \BibitemOpen
  \bibfield  {author} {\bibinfo {author} {\bibfnamefont {I.~M.}\ \bibnamefont {Pop}}, \bibinfo {author} {\bibfnamefont {I.}~\bibnamefont {Protopopov}}, \bibinfo {author} {\bibfnamefont {F.}~\bibnamefont {Lecocq}}, \bibinfo {author} {\bibfnamefont {Z.}~\bibnamefont {Peng}}, \bibinfo {author} {\bibfnamefont {B.}~\bibnamefont {Pannetier}}, \bibinfo {author} {\bibfnamefont {O.}~\bibnamefont {Buisson}},\ and\ \bibinfo {author} {\bibfnamefont {W.}~\bibnamefont {Guichard}},\ }\bibfield  {title} {\bibinfo {title} {Measurement of the effect of quantum phase slips in a josephson junction chain},\ }\href {https://doi.org/10.1038/nphys1697} {\bibfield  {journal} {\bibinfo  {journal} {Nature Physics}\ }\textbf {\bibinfo {volume} {6}},\ \bibinfo {pages} {589} (\bibinfo {year} {2010})}\BibitemShut {NoStop}%
\end{thebibliography}%

\end{document}